\RequirePackage{fix-cm}
\documentclass[twocolumn,epjc3]{svjour3}  
\smartqed  
\RequirePackage{graphicx}
\RequirePackage{latexsym}
\RequirePackage[numbers,sort&compress]{natbib}
\RequirePackage[colorlinks,citecolor=blue,urlcolor=blue,linkcolor=blue]{hyperref}
\RequirePackage{amsmath,amssymb}
\def\beq{\begin{equation}}
\def\eeq{\end{equation}}
\def\beqn{\begin{eqnarray}}
\def\eeqn{\end{eqnarray}}

\journalname{Eur. Phys. J. A}
\begin{document}

\title{The effect of the energy functional on the pasta-phase properties of catalysed neutron stars 
}

\titlerunning{The effect of the energy functional on the pasta-phase properties}        

\author{H. Dinh Thi\thanksref{e1,addr1}
		\and
		A. F. Fantina\thanksref{e2,addr2}
        \and
        F. Gulminelli\thanksref{e3,addr1} 
}

\thankstext{e1}{e-mail: dinh@lpccaen.in2p3.fr}
\thankstext{e2}{e-mail: anthea.fantina@ganil.fr}
\thankstext{e3}{e-mail: gulminelli@lpccaen.in2p3.fr}


\institute{Normandie Univ., ENSICAEN, UNICAEN, CNRS/IN2P3, LPC Caen, F-14000 Caen, France \label{addr1}
           \and
           Grand Acc\'el\'erateur National d’Ions Lourds (GANIL), CEA/DRF - CNRS/IN2P3, Boulevard Henri Becquerel, 14076
Caen, France \label{addr2}
}

\date{Received: date / Accepted: date}

\maketitle

\begin{abstract}
Nuclear pasta, that is an inhomogeneous distribution of nuclear matter characterised by non-spherical clustered structures, is expected to occur in a narrow spatial region at the bottom of the inner crust of neutron stars, but the width of the pasta layer is strongly model dependent. 
In the framework of a compressible liquid-drop model, we use Bayesian inference to analyse the constraints on the sub-saturation energy functional and surface tension imposed by both ab-initio chiral perturbation theory calculations and experimental measurements of nuclear masses. 
The posterior models are used to obtain general predictions for the crust-pasta and pasta-core transition with controlled uncertainties. 
A correlation study allows extracting the most influential parameters for the calculation of the pasta phases. 
The important role of high-order empirical parameters and the surface tension is underlined.

\end{abstract}

%
%

\section{Introduction}
\label{sec:intro}

Born hot from core-collapse supernova explosions, neutron stars (NSs) are generally assumed to cool down until the ground state at zero temperature is eventually reached.
We will consider here NSs under this so-called ``cold catalysed matter'' hypothesis, meaning that full (thermodynamic, nuclear, and beta) equilibrium holds and that the temperature is low enough that thermal effects can be neglected for the equation of state and composition.
At least three regions can be identified in such NSs: the outer crust, made of ions arranged in a regular lattice, embedded in an electron gas, the inner crust, where the neutron-proton clusters, neutralised by the electron gas, also coexist with a neutron gas, and a core, starting from about half the saturation density ($n_{\rm sat} \approx 0.15$~fm$^{-3}$), where nuclear clusters have dissolved into homogeneous matter \cite{hpy2007, Blaschke2018}.
At the lower densities in the crust, the internal ion structure is not influenced by neighbouring ions and their geometry is thus expected to be spherical.
However, in the innermost layers of the inner crust, the energetically favoured configuration might rather be that of a spatially periodic distribution of inhomogeneities with non-spherical symmetry, collectively known as ``pasta'' phases.

From the observational point of view, so far there is no direct evidence of the existence of pasta phases.
However, their presence may have a considerable impact on different NS phenomena related to the NS crust size, such as magnetic, thermal, and rotational evolution of NSs \cite{Pons2013, Vigano2013}, NS cooling \cite{Newton2013b, hor2015, lin2020}, transport properties (see, e.g., \cite{SchSht2018} for a review), crust oscillations \cite{gearheart2011, sotani2012}, and NS ellipticity that can be potentially measured by gravitational-wave observations \cite{gearheart2011}.

From the theoretical point of view, while the existence of these exotic structures has been predicted in the NS inner crust since the '80s \cite{Ravenhall1983, Hashimoto1984, oya1984}, their actual presence and properties still remain model dependent.
Several investigations on the nuclear pasta have been performed so far, within different theoretical frameworks.
These latter include compressible liquid-drop (CLD) models (e.g., \cite{LRP1993, Newton2013, Pais2016, lh2017, Vinas2017, Balliet2020} for recent works), (extended) Thomas-Fermi method (e.g., \cite{Sharma2015, lh2017, Martin2015, Vinas2017, Pearson2020, Sengo2020, Ji2021}), nuclear energy density functional theory (e.g., \cite{Schu2015, Sagert2016, Fattoyev2017}), and molecular dynamics calculations (e.g. \cite{hor2015, Sch2016, Berry2016, lin2020}); for a review, see also refs.~\cite{Pethick1995, ch2008, wm2012, maru2012, Blaschke2018} and references therein.
In particular, some of these works found that a sizeable amount of the crust could be made up by pasta \cite{LRP1993, Newton2013, Balliet2020}.
Specifically, in the recent work by Balliet et al.~\cite{Balliet2020}, where a Bayesian analysis has been employed to calculate the distributions of pasta observables, it was found that the pasta contributes for more than 50\% of the crust by mass and 15\% by thickness.

Very recently, we have extended the CLD model of Carreau et al.~\cite{Carreau2019, Carreau2020} to account for non-spherical pasta structures in the inner crust \cite{dinh2021}. 
Our outcomes also show that the presence of pasta phases is robustly predicted in an important fraction of the inner crust.
Moreover, we have highlighted the importance of a consistent calculation of the nuclear functional, both the bulk and finite-size contributions, and found that the surface and curvature parameters are influential for the description of the pasta observables. 
In this work, we pursue the study of ref.~\cite{dinh2021} on the properties of the pasta phases in cold catalysed nonaccreting NSs, with a particular focus on the nuclear functional. Specifically, we use Bayesian inference to extract the distribution of the nuclear matter empirical parameters and of the isospin-dependent surface and curvature energy. 
The posterior parameter distributions obtained from the constraint of nuclear mass measurements, and chiral perturbation theory calculation of the low-density equation of state, are used to pin down the most influential nuclear parameters for the calculation of the pasta phase, and to obtain general predictions for the transition between the solid inner crust and the amorphous pasta layer, and the transition from the pasta to the homogeneous liquid core.

We recall the formalism in Sect.~\ref{sec:model} and present the numerical results in Sect.~\ref{sec:res}.
In particular, the general prediction of the crust-pasta and pasta-core transition, as obtained from the different theoretical and experimental constraints, is shown in Sect.~\ref{sec:bayes}.
 We discuss the posterior distribution of the bulk and surface empirical parameters in Sect.~\ref{sec:bulk} and Sect.~\ref{sec:surf} respectively, while in Sect.~\ref{sec:corr} we examine the influence of the different parameters on the pasta transitions through a correlation study.
Finally, we present our conclusions in Sect.~\ref{sec:concl}.

\section{Theoretical framework}
\label{sec:model}

We model the pasta phase at the bottom of the inner crust as described in \cite{dinh2021}.
We recall here the main points and assumptions.

We consider the inner crust of a catalysed NS composed of a periodic lattice consisting of Wigner-Seitz cells of volume $V_{\rm WS}$ containing a clustered structure (``pasta'') with $Z$ protons of mass $m_p$ and $A-Z$ neutrons of mass $m_n$, $A$ being the cluster total mass number.
The surrounding uniform gas of neutrons and electrons has densities $n_g$ and $n_e$, respectively.
If inhomogeneities are clusters (holes) of volume $V$, the density distribution in the Wigner-Seitz cell is $n_i$ ($n_g$) if $l<r_N$, and $n_g$ ($n_i$) otherwise, $r_N$ being the linear dimension of the pasta structure and $l$ the linear coordinate of the cell.
The density of the denser phase is $n_i=A/V$ in the case of cluster ($n_i=A/(V_{\rm WS}-V)$ in the case of holes) and its proton fraction is given by $y_p=Z/A=n_p/n_i$. 
The volume fraction occupied by the cluster (of density $n_i$) or hole (of density $n_g$), $u=V/V_{\rm WS}$, is thus given as
\begin{equation}
u=  \left \{
   \begin{array}{r c l}
      (n_B-n_g)/(n_i-n_g)  &  & \mbox{for clusters}, \\
     (n_i-n_B)/(n_i-n_g)  &  & \mbox{for holes}.
    \end{array}
   \right . 
\end{equation}

Since we assume that matter be at zero temperature, the NS crust is in its absolute ground state.
Therefore, its equilibrium configuration is obtained variationally, by minimising the energy density of the Wigner-Seitz cell under the constraint of a given baryon density $n_B=n_n +n_p$, $n_p$ ($n_n$) being the proton (neutron) density, respectively (see, e.g., the seminal work by Baym et al.~\cite{bbp}).
Since charge neutrality holds, the electron and proton densities are equal, i.e. $n_e=n_p$.
In this work, we employ a CLD model approach, along the lines of refs.~\cite{Douchin2001, Carreau2019, Carreau2020}.
The thermodynamic potential per unit volume of the cell can thus be written as: 
\begin{eqnarray}
\Omega
&=& n_n m_n c^2+n_p m_p c^2 + \epsilon_B(n_i,1-2y_p) f(u) \nonumber \\ 
&+& \epsilon_B(n_g,1) (1-f(u)) + \epsilon_{\rm Coul} + \epsilon_{\rm {surf+curv}} \nonumber \\
&+& \epsilon_e -\mu_B^{\rm tot} \ n_B , 
\label{eq:auxiliary}
\end{eqnarray}
where $\epsilon_B(n,\delta)$ is the energy density of uniform nuclear matter at density $n$ and isospin asymmetry $\delta=(n_n-n_p)/n$, $\epsilon_e$ is the energy density of a pure uniform electron gas, $\mu_B^{\rm tot}$ is the baryonic chemical potential (including the rest mass), and the finite-size terms $\epsilon_{\rm {surf+curv}}$ and $\epsilon_{\rm Coul}$ account for the interface tension between the cluster and the neutron gas, and the electrostatic energy density, respectively. 
The function $f(u)$ appearing in Eq.~(\ref{eq:auxiliary}) reads
\begin{equation}
f(u)=  \left \{
   \begin{array}{r c l}
      u  &  & \mbox{for clusters}, \\
     1-u  &  & \mbox{for holes}.
    \end{array}
   \right . 
\end{equation}
One has then to choose a nuclear model, i.e. the energy functional $\epsilon_B(n,\delta)$, supplemented with the interface energy density, $\epsilon_{\rm {surf+curv}}$, that we discuss in the next sections.

\subsection{The nuclear functional}
\label{sec:nucfunc}

For the nuclear functional, we employ the meta-modelling approach of refs.~\cite{Margueron2018a, Margueron2018b}.
To this aim, we introduce a Taylor expansion in $x=(n-n_{\rm sat})/3n_{\rm sat}$ up to order $N$ around the saturation point $(n=n_{\rm sat},\delta=0)$, where the parameters of the expansion correspond to the so-called equation-of-state empirical parameters \cite{Piekar2009}:
\begin{equation}
\epsilon_B(n,\delta)\approx n \sum_{k=0}^N \frac{1}{k!} \left ( \left. \frac{d^k e_{\rm sat}}{d x^k} \right|_{x=0} 
+ \left.  \frac{d^k e_{\rm sym}}{d x^k} \right|_{x=0}\delta^2\right ) x^k .
\label{eq:etotisiv}
\end{equation} 
In Eq.~(\ref{eq:etotisiv}), $e_{\rm sat}=\epsilon_B(n,0) / n$ is the energy per baryon of symmetric matter and $e_{\rm sym}=(\epsilon_B(n,1)-\epsilon_B(n,0))/ n$ is the symmetry energy per baryon, which is defined here as the difference between the energy of pure neutron matter and that of symmetric matter. 
In ref.~\cite{Margueron2018a}, it has been shown that it is possible to explore the different behaviours of the functional truncating the expansion at order $N=4$, provided that: 
(i) the dominant Fermi gas $n^{5/3}$ term and the associated correction to the parabolic approximation with a $\delta^{5/3}$ term is added, 
(ii) different values for the coefficients of order 3 and 4 are used in the sub-saturation and supra-saturation regime, 
(iii) an exponential correction is introduced ensuring the correct limiting behaviour at zero density. 
This yields the following final expression for the energy density in the meta-modelling approach (see Eq.~(17) in ref.~\cite{Carreau2019}):
\begin{eqnarray}
\epsilon_B(n,\delta) &=& n \frac{3 \hbar^2}{20m} \left( \frac{3 \pi^2 n}{2} \right)^{2/3} \left[ \left( 1 + \mathcal{K}_{\rm sat} \frac{n}{n_{\rm sat}} \right) f_1  \right. \nonumber \\
&& \left. +\mathcal{K}_{\rm sym} \frac{n}{n_{\rm sat}} f_2 \right] \nonumber \\
 &+ & n \sum_{k=0}^{N=4} \frac{1}{k!} \left ( 
 X^{(k)}_{\rm sat} 
+ 
X^{(k)}_{\rm sym}\delta^2\right ) x^k \nonumber \\
&& - (a_N^{\rm is} + a_N^{\rm iv} \delta^2) \ x^{N+1} \exp\left(-b \frac{n}{n_{\rm sat}}\right) \ ,
\label{eq:etot}
\end{eqnarray}
with $m$ the nucleon mass, the parameters $a_N^{\rm is}$ and $a_N^{\rm iv}$ being entirely fixed by the condition at zero density in symmetric matter and neutron matter, and 
\begin{eqnarray}
f_1(\delta) &=& (1+\delta)^{5/3} + (1-\delta)^{5/3} \\
f_2(\delta) &=& \delta \left[ (1+\delta)^{5/3} - (1-\delta)^{5/3} \right] \ .
\end{eqnarray}
Following the common notation in the literature, we denote $E_{\rm sat(sym)}=e_{\rm sat(sym)}(n=n_{\rm sat})=X^{(0)}_{\rm sat(sym)}$, while the other parameters $X^{(k)}_q$ corresponding to the successive derivatives of $e_q$, with $q=$~sat,sym, are called $L_q,K_q,Q_q,Z_q$. 
The bulk parameters \\
$\{E_q,L_q,K_q,Q_q,Z_q, q=$~sat,sym$\}$ \\
sare complemented with the saturation density parameter, $n_{\rm sat}$, the two parameters related to the isoscalar effective mass at symmetric matter saturation $\mathcal{K}_{\rm sat} = m/m^\star_{\rm sat} - 1$ and effective mass splitting $\mathcal{K}_{\rm sym} = (m/m^\star_n - m/m^\star_p)/2 $, and the $b$ parameter governing the functional behaviour close to the zero-density limit (see Sect.~2.2 in ref.~\cite{Carreau2019} for details).
The complete parameter set describing the sub-saturation equation of state  thus has 13 parameters and will be noted in a compact form as \\
$\vec X\equiv\{n_{\rm sat}, b, (E_q,L_q,K_q,Q_q,Z_q, \mathcal{K}_q, q=$~sat,sym$)\}$.\\
Different nuclear models will then correspond to different sets of $\vec X$ parameters.

\subsection{The finite-size contributions}
\label{sec:finsize}

To model the inhomogeneities in the inner crust, the bulk term in the energy density has to be supplemented with the finite-size contributions, namely the interface and Coulomb energy density, $\epsilon_{\rm {surf+curv}}$ and $\epsilon_{\rm {Coul}}$ in Eq.~(\ref{eq:auxiliary}).
An advantage of such a decomposition in Eq.~(\ref{eq:auxiliary}) is that the geometry of the pasta structures only enters in the finite-size terms (see, e.g., the pioneer works of refs.~\cite{Ravenhall1983,Hashimoto1984}), which in turn can be expressed as a function of the structure dimensionality $d$ ($d=1$ for slabs, $d=2$ for cylinders, $d=3$ for spheres).

For the interface energy density, we employ the same expression as in refs.~\cite{Maru2005,Newton2013},
\begin{equation}
\epsilon_{\rm {surf+curv}}=\frac{ud}{r_N}\left ( \sigma_s +\frac{(d-1)\sigma_c}{r_N}\right ) , \label{eq:interface}   
\end{equation}
where $\sigma_s$ is the surface tension and $\sigma_c$ is the curvature tension, both independent of the dimensionality.
We adopt here the expressions of $\sigma_s$ and $\sigma_c$ as originally proposed in ref.~\cite{Ravenhall1983}, based on Thomas-Fermi calculations at extreme isospin asymmetries:
\begin{eqnarray}
\sigma_s&=&\sigma_0\frac{2^{p+1}+b_s}{y_p^{-p}+b_s+(1-y_p)^{-p}} \ , \label{eq:surface} \\
\sigma_c&=&5.5 \, \sigma_s \frac{\sigma_{0,c}}{\sigma_0}(\beta-y_p)\ , \label{eq:curvature} \ 
\end{eqnarray}
where the parameters $(\sigma_0,\sigma_{0,c},b_s,\beta,p)$ must be optimised on theoretical calculations or experimental data. 
 
The Coulomb energy density reads:
\begin{equation}
\epsilon_{\rm Coul}= 2\pi \left ( e y_p n_i r_N \right )^2 u \eta_d , \label{eq:coulomb}   
\end{equation}
where $e$ is the elementary charge and the function $\eta_d(u)$ is given by
\begin{eqnarray}
\eta_1&=& \frac{1}{3}\left [ u- 2 \left ( 1- \frac{1}{2u}\right )  \right ],  \label{eq:eta1} \\
\eta_2&=& \frac{1}{4}\left [ u - \ln u -1\right ],  \label{eq:eta2} \\
\eta_3&=&\frac{1}{5}\left [ u+ 2 \left ( 1- \frac{3}{2}u^{1/3} \right ) \right ]. \label{eq:eta3}
\end{eqnarray}

In the vacuum, the nuclear mass of a spherical fully ionised atom of charge $Z$ and mass number $A$ can be deduced using Eqs.~(\ref{eq:auxiliary}), (\ref{eq:interface}), and (\ref{eq:coulomb}) as:
\begin{eqnarray}
M(A,Z) c^2 &=& m_pc^2 Z+m_nc^2( A-Z) \nonumber \\ 
&+& \frac{A}{n_0}  \epsilon_B(n_0,I) +4\pi r_N^2\left (\sigma_s +\frac{2\sigma_c}{r_N}\right ) \nonumber \\ 
&+& \frac{3}{5}\frac{e^2Z^2}{r_N} , \label{eq:mass}
\end{eqnarray} 
where $I=1-2Z/A$, the (spherical) nuclear radius is $r_N=(3/4\pi n_0)^{1/3}A^{1/3}$, and the bulk density $n_0$ is given by the equilibrium density of nuclear matter at isospin asymmetry $I$, defined by $\partial \epsilon_B/\partial n|_{I,n_0}=0$.

In previous applications of Eq.~(\ref{eq:interface}) on the NS crust and supernova modelling within the CLD approximation \cite{lattimer1991,LRP1993,Newton2013,lh2017,Lim2019a,Lim2019b,Balliet2020} the surface parameters were fixed on Thomas-Fermi or Hartree-Fock calculations, independently of the bulk functional. 
However, both bulk and surface terms must be specified to variationally obtain the matter composition, and they are clearly correlated notably by the constraint of reproducing the nuclear mass, which is experimentally known for a large panel of nuclei in the vacuum. 
For this reason, in this work we include the uncertainty on the surface energy by adding  $(\sigma_0,\sigma_{0,c},b_s,\beta,p)$ to our parameter space. 
Following refs.~\cite{Carreau2019, Carreau2020}, for each choice of the parameter set $\vec X$, the associated surface parameters are determined by a $\chi^2$-fit of Eq.~(\ref{eq:mass}) to the experimental Atomic Mass Evaluation (AME) 2016 \cite{AME2016}. 
Well defined minima are found for all surface parameters but $p$, which governs the behaviour of the surface tension at extreme isospin values. 
This latter is added as an extra independent parameter to our set $\vec X$.

With the choice for the energy functional $\epsilon_B(n,\delta)$, described in Sect.~\ref{sec:bulk}, and for a set of values for the parameters $\{\vec X, p\}$, the nuclear model is thus specified.
The pasta structure and composition at a given baryonic density $n_B$ in the crust are then determined by a two-steps process: 
(i) a geometry (with dimensionality $d$) and a shape (clusters or holes) are considered, and 
(ii) the thermodynamical potential Eq.~(\ref{eq:auxiliary}) is minimised with respect to the variational parameters $(n_i,I=1-2y_p, A,n_p,n_g)$. 
This allows to identify the baryonic chemical potential $\mu = \mu_B^{\rm tot} - m_n c^2$ with the chemical potential of the neutron gas,
\begin{equation}
\mu=\frac{d\epsilon_B(n_g,1)}{dn_g} \  ,
\end{equation}
and to find the optimal value of the thermodynamic potential for each geometry, i.e. $\Omega=\Omega_{\rm opt}$. 
The equilibrium configuration thus corresponds to the geometry (or, equivalently, the dimensionality $d$) and the shape (cluster or hole) for which the minimum value of $\Omega_{\rm opt}$ is obtained.

\section{Pasta-phase properties}
\label{sec:res}

\begin{figure}
  \includegraphics[width=\linewidth]{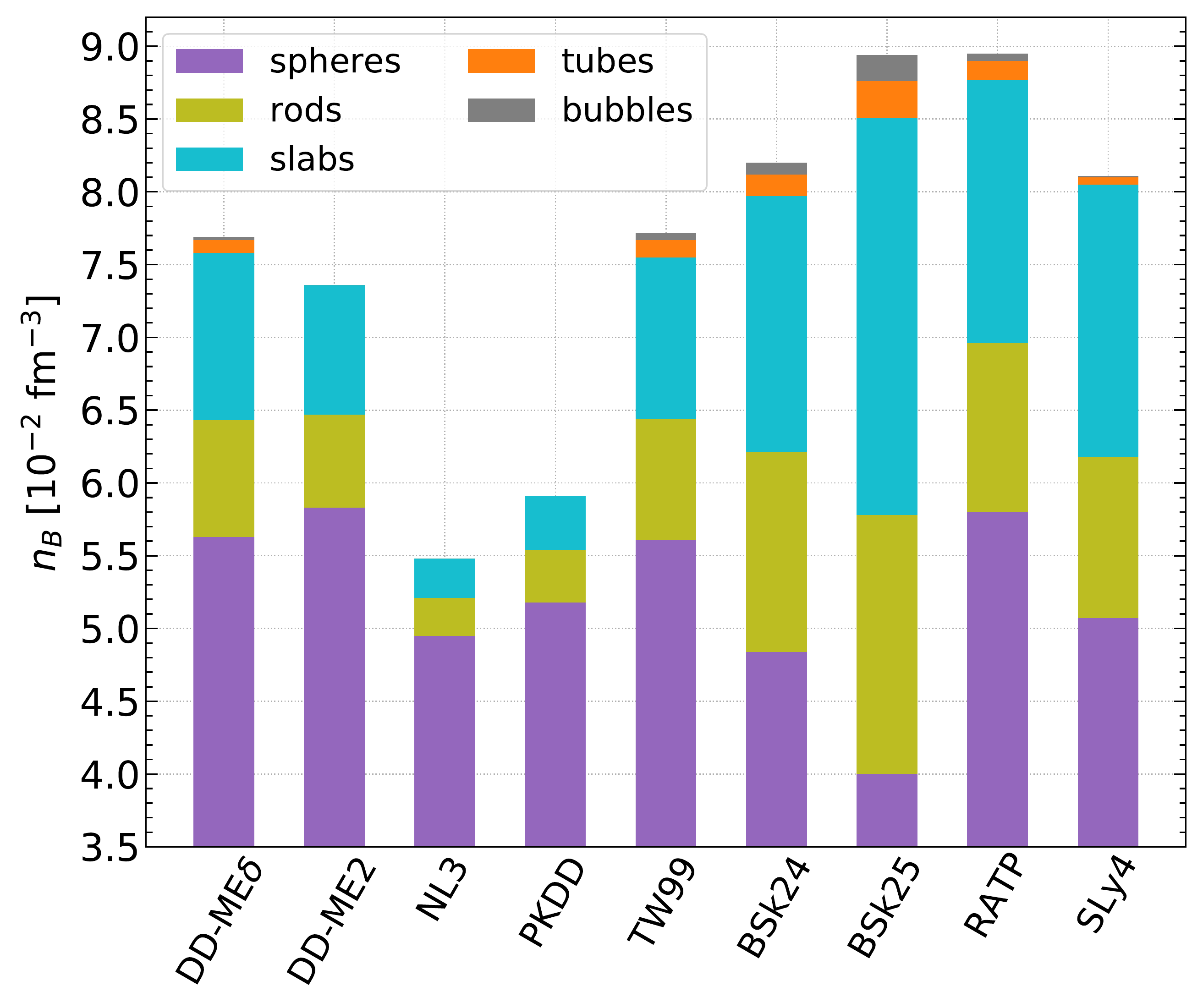}
\caption{Sequence of pasta phases at the bottom of the inner crust and corresponding transition densities for different nuclear functionals. Colours correspond to the different geometries. See text for details.}
\label{fig:models}       
\end{figure}

\begin{figure*}
\centering
  \includegraphics[width=0.75\linewidth]{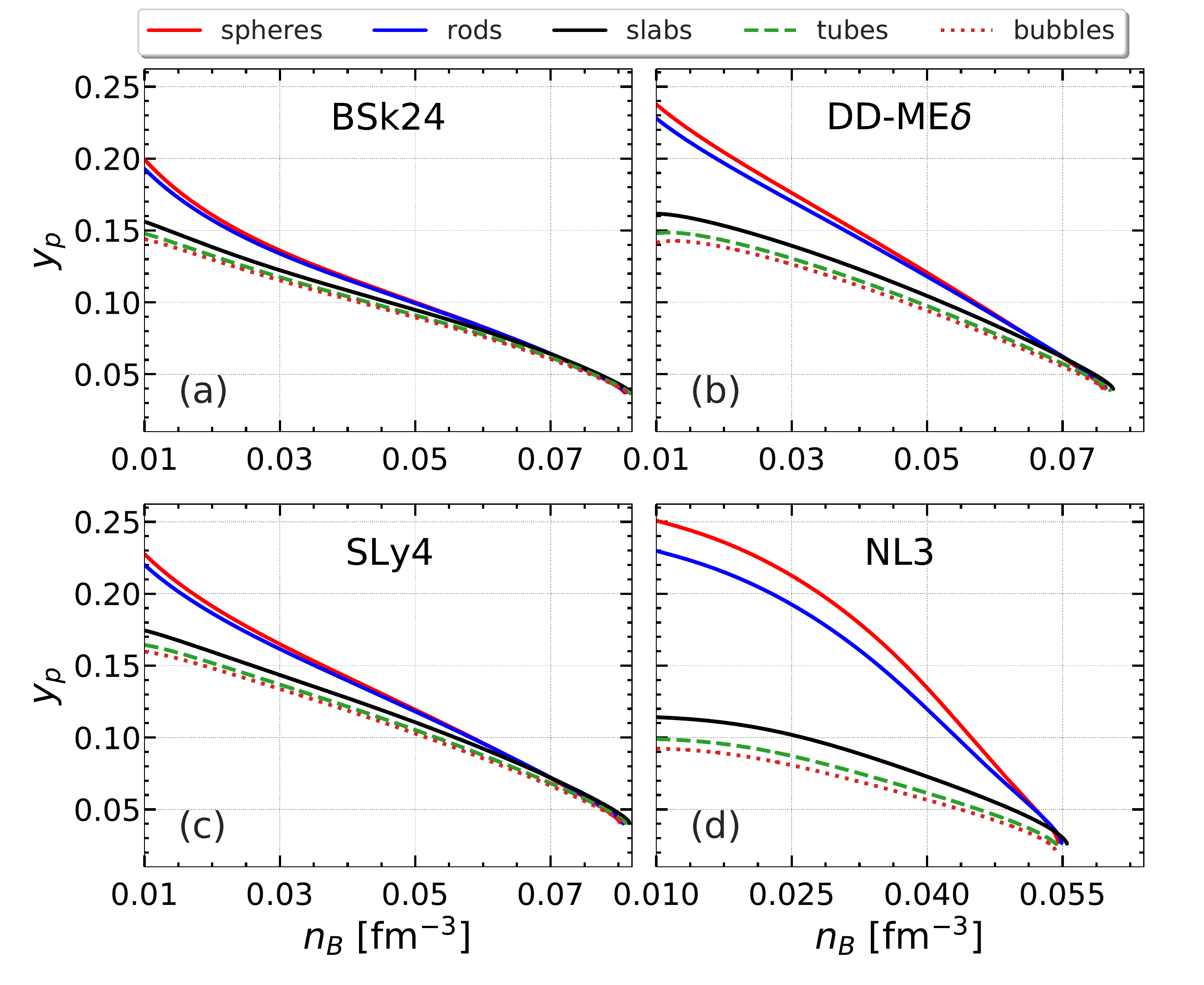}
\caption{Proton fraction of the clustered structure, $y_p=Z/A$, as a function of the baryonic density in the neutron-star crust for different geometries using different nuclear models. See text for details.}
\label{fig:yp}       
\end{figure*}

Employing the model described in Sect.~\ref{sec:model}, we have calculated the properties of the pasta phase which is predicted to appear at the bottom of the inner crust. 
We start the discussion by showing in Fig.~\ref{fig:models} the sequence of the equilibrium configurations obtained with different nuclear meta-models, i.e. with different parameter sets $\mathbf{X}$ corresponding to the non-relativistic functionals BSk24 and BSk25 \cite{BSK24}, SLy4 \cite{SLy4} and RATP \cite{RATP}, and the relativistic functionals DD-ME2 \cite{DDME2}, DD-ME$\delta$ \cite{DDMEd}, NL3 \cite{NL3}, PKDD \cite{PKDD}, and TW99 \cite{TW99}.
For these calculations, the $p$ parameter has been optimised to provide a good reproduction of the crust-core transition density of the different functionals, whenever available, or fixed to $p=3$ otherwise (see Table~2 in ref.~\cite{dinh2021} and refs.~\cite{Carreau2019, Carreau2020} for a discussion).
The different colours represent the density ranges where spheres, rods, slabs, tubes, and eventually bubbles dominate.
We can see that, while the transition densities are model dependent, the sequence still remains the same for the different models, although not all functionals predict the existence of bubbles.
As already noticed in ref.~\cite{dinh2021}, these results are in good agreement with those available in the literature for the transition densities from sphere to cylinders (rods), for the functionals BSk24 \cite{Pearson2020}, SLy4 \cite{Martin2015, Douchin2001, Vinas2017}, NL3, DD-ME2, and DD-ME$\delta$ \cite{Grill2012}.
Note that we adopt a different fitting protocol for the finite-size parameters with respect to the aforementioned works; therefore, the agreement can be considered very satisfactory. 

In Fig.~\ref{fig:yp} we show the proton fraction of the clustered (pasta) structure, $y_p=Z/A$, as a function of the baryon density for some selected models, namely the non-relativistic (meta-)models BSk24 \cite{BSK24} (panel (a)) and SLy4 \cite{SLy4} (panel (c)), and the relativistic (meta-)models DD-ME$\delta$ \cite{DDMEd} (panel (b)) and NL3 \cite{NL3} (panel (d)), as illustrative examples.
We can see that the values of $y_p$ for the considered models do not exceed $0.25$, and in the high-density regime corresponding to the pasta phase the typical proton fraction of the clustered structure varies between $0.05$ and $0.1$.
This underlines the importance of determining the surface tension at extreme isospin values. 
We can also observe that the proton fraction for non-spherical configurations is systematically lower than that of the spheres, which follows the general trend of the geometry dependence of the global proton fraction in the Wigner-Seitz cell (see Fig.~2 in ref.~\cite{dinh2021}).

From Figs.~\ref{fig:models} and \ref{fig:yp} we can observe that the predictions for the pasta properties clearly exhibit some model dependence.
In order to quantitatively address this issue and determine the influence of the choice of the functionals on the uncertainties on the pasta observables, we have performed a Bayesian analysis, which is discussed in the next section.

\subsection{Statistical analysis}
\label{sec:bayes}

\begin{figure}
\centering
  \includegraphics[width=0.75\linewidth]{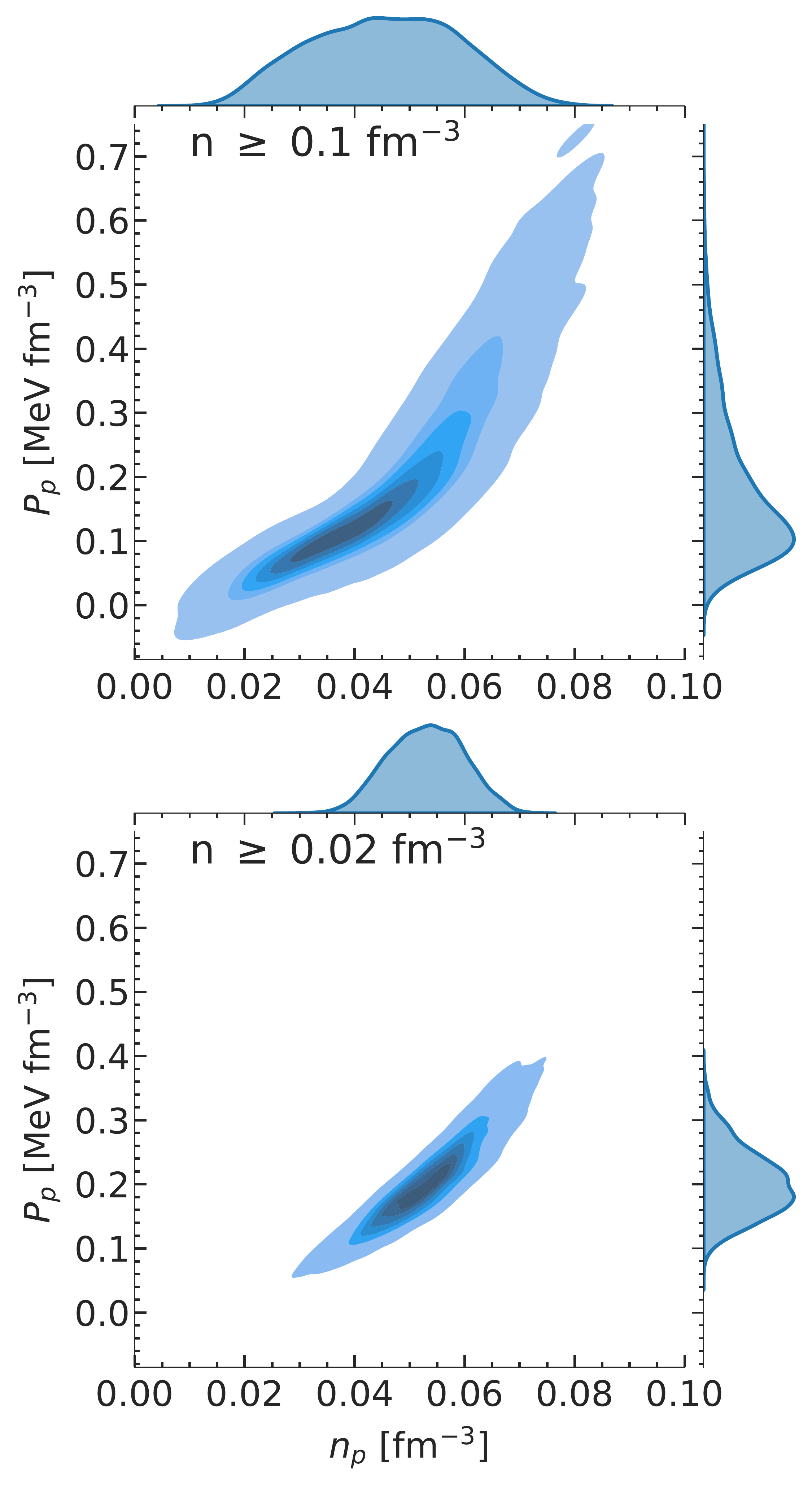}
\caption{Posterior correlation between the pressure and the density at the interface between the inner crust and the emergence of non-spherical pasta structures. In the upper panel the low-density filter is applied from 0.1~fm$^{-3}$ to 0.2~fm$^{-3}$, while in the lower panel it is applied from 0.02~fm$^{-3}$ to 0.2~fm$^{-3}$. }
\label{fig:ppasta_npasta}       
\end{figure}

\begin{figure}
\centering
  \includegraphics[width=0.75\linewidth]{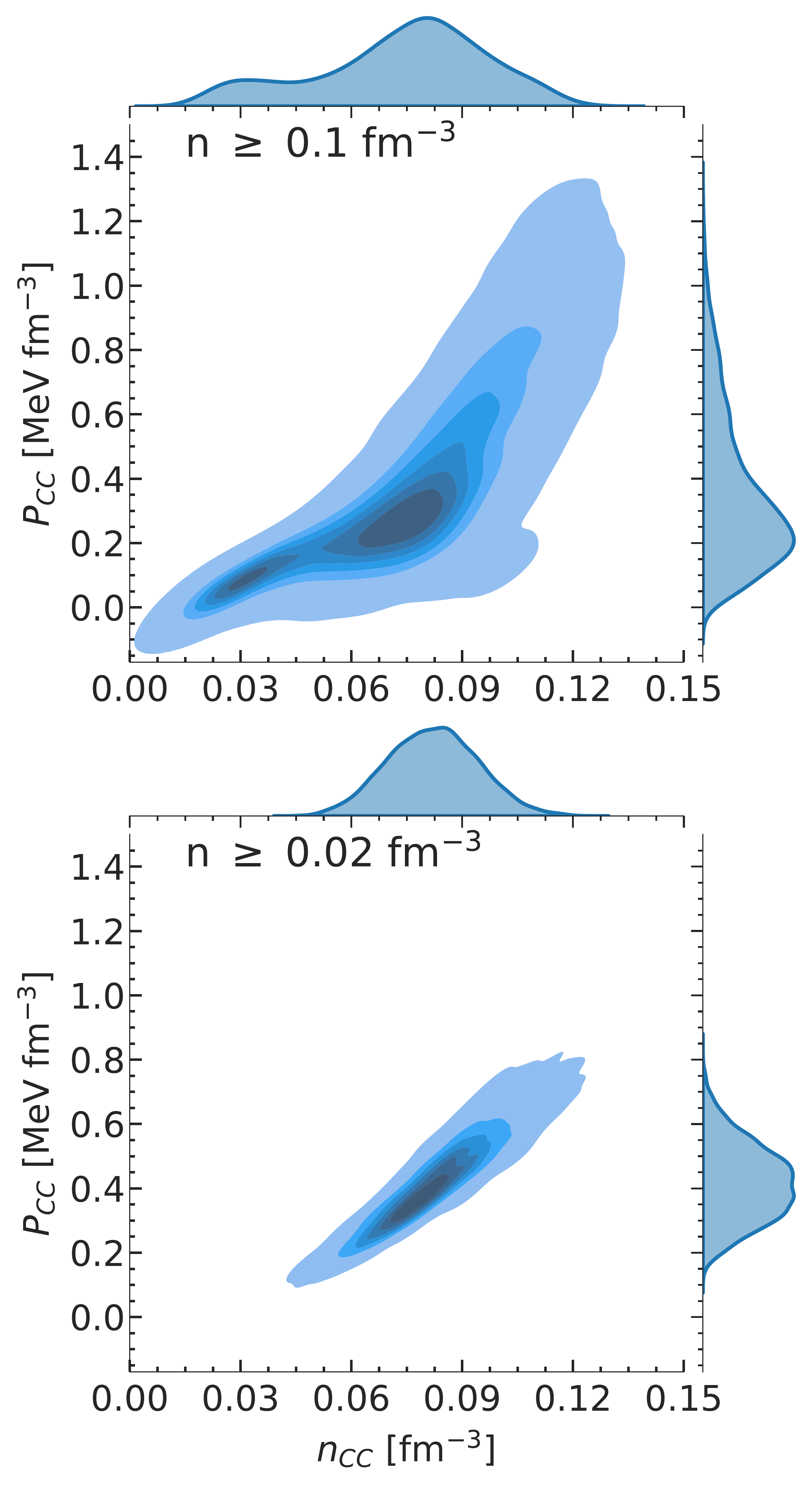}
\caption{Posterior correlation between the pressure and the density at the interface between the pasta phase and the homogeneous solution. In the upper panel the low-density filter is applied from 0.1~fm$^{-3}$ to 0.2~fm$^{-3}$, while in the lower panel it is applied from 0.02~fm$^{-3}$ to 0.2~fm$^{-3}$. }
\label{fig:pcc_ncc}       
\end{figure}

\begin{table}
\caption{Minimum and maximum values of the parameter set $\mathbf{X}$.}
\label{tab:param-minmax}      
\centering
\begin{tabular}{lll}
\hline\noalign{\smallskip}
Parameter & Min & Max  \\
\noalign{\smallskip}\hline\noalign{\smallskip}
$E_{\rm sat}$ [MeV] & -17 & -15 \\
$n_{\rm sat}$ [fm$^{-3}$] & 0.15 & 0.17 \\
$K_{\rm sat}$ [MeV] & 190 & 270 \\
$Q_{\rm sat}$ [MeV] & -1000 & 1000 \\
$Z_{\rm sat}$ [MeV] & -3000 & 3000 \\
$E_{\rm sym}$ [MeV] & 26 & 38 \\
$L_{\rm sym}$ [MeV] & 10 & 80 \\
$K_{\rm sym}$ [MeV] & -400 & 200 \\
$Q_{\rm sym}$ [MeV] & -2000 & 2000 \\
$Z_{\rm sym}$ [MeV] & -5000 & 5000 \\
$m^\star_{\rm sat}/m$ & 0.6 & 0.8 \\
$\Delta m^\star_{\rm sat}/m$ & 0.0 & 0.2 \\
$b$ & 1 & 6 \\
$p$ & 2 & 4 \\
\noalign{\smallskip}\hline
\end{tabular}
\end{table}

Starting from flat non-informative priors obtained by largely varying the model parameters $\vec X$ \footnote{In this section and the following ones, the surface parameter $p$ is systematically included in the $\vec X$ parameter set as an additional independent parameter, with a flat prior distribution as given in Table \ref{tab:param-minmax}.}, we apply both low-density (LD) constraints accounting for our present knowledge of nuclear physics and high-density (HD) constraints coming from general and NS physics.
Both strict filters ($w_{\rm LD(\rm HD)}$), and likelihood expressions ($w_{\rm mass}$) are applied to the prior distribution to generate the posterior distribution
\begin{eqnarray}
p_{\rm post} (\vec X) = \mathcal{N} \, w_{\rm LD}(\vec X) w_{\rm HD}(\vec X) \, w_{\rm mass}(\vec X) \, p_{\rm prior}(\vec X)  ,
\label{eq:probalikely}
\end{eqnarray}
where $\mathcal{N}$ is the normalization.
The $w_{\rm LD}$ filter is given by the uncertainty band of the chiral N$^3$LO effective field theory (EFT) calculations of the energy per particle of symmetric and pure neutron matter by Drischler et al.~\cite{Drischler2016}, which is interpreted as a $90\%$ confidence interval.
Since the EFT energy band becomes very small at low density, in ref.~\cite{Carreau2019} the $w_{\rm LD}$ filter was applied from $\approx 0.1$~fm$^{-3}$, while in ref.~\cite{dinh2021} we have extended this constraint to lower densities, namely in the range $[0.02-0.2]$~fm$^{-3}$.
We further investigate the effect of such a choice in Sect.~\ref{sec:bulk}.

The $w_{\rm HD}$ filter is defined by imposing 
(i) stability, i.e. the derivative of the pressure with respect to the mass-energy density must be positive, $dP/d\rho \ge 0$, 
(ii) causality, i.e. the speed of sound must be positive and smaller than the speed of light, 
(iii) a positive symmetry energy at all densities, and 
(iv) the resulting equation of state to support $M_{\rm max} > 1.97M_\odot$,  where $M_{\rm max}$ is the maximum NS mass at equilibrium determined from the solution of the Tolmann-Oppenheimer-Volkoff (TOV) equations \cite{hpy2007} ($M_\odot$ being the solar mass).
Finally, $w_{\rm mass}$ quantifies the quality of experimental nuclear mass reproduction of each $\vec X$ set,  
\begin{equation}
w_{\rm mass}(\vec X) = \exp \left[-\sum_n \frac{\left ( M(A_n,Z_n)-M_{\rm exp}(A_n,Z_n)\right )^2}{2 s^2} \right] \ ,
\label{eq:chi2}
\end{equation}
where the sum runs over the AME2016 nuclear mass table \cite{AME2016}, $M$ is calculated from Eq.~(\ref{eq:mass}) for each model, $s$ corresponds to the average systematic theoretical error, and the parameters ($\sigma_0,\sigma_{0,c},b_s,\beta$) are the ones that maximise Eq.~(\ref{eq:chi2}) for the model set $\vec X$ under study. 
Moreover, an additional constraint is given by the condition that the minimisation of the thermodynamic potential Eq.~(\ref{eq:auxiliary}) leads to physically meaningful results for the crust, namely positive values for the gas and cluster densities.
The equation of state thus obtained was shown to be compatible with the measurement of the tidal polarizability extracted from the gravitational-wave event GW170817 \cite{Abbott2018} (see refs.~\cite{Carreau_prc, dinh2021} for details).

We have generated $10^8$ models to numerically sample the prior parameter distribution; of those, 7008 models are retained when the low-density EFT filter is applied from 0.02~fm$^{-3}$.
In order to have comparable statistics, $2 \times 10^6$ models are generated, of which 7714 are retained, when the EFT filter is applied from 0.1~fm$^{-3}$. 
From the marginalized posteriors, the average value of the different observables $Y$ is thus obtained as
\begin{equation}
\langle Y \rangle = \prod_{k=1}^{14} \int_{X_k^{\rm min}}^{X_k^{\rm max}} dX_k Y(\vec X) p_{\rm post}(\vec X),
\label{eq:distri}
\end{equation}
where $p_{\rm post}(\vec X)$ is the posterior distribution, $Y(\vec X)$ is the value of the $Y$ variable as obtained with the $\vec X$ parameter set, $X_k^{\rm min(max)}$ is the minimum (maximum) value in the prior distribution of parameter $X_k$.
The latter values for the different $\vec X$ parameters are given in Table~\ref{tab:param-minmax} (see also refs.~\cite{Carreau_these,Margueron2018a} for details). 
Moreover, to speed up the computation in the Bayesian analysis, the composition of the different phases are fixed to those found for the spheres. 
Indeed, we have verified that this choice has a negligible impact on the sphere-pasta transition point.

The posterior distribution of the pressure and density at the crust-pasta interface is displayed in Fig.~\ref{fig:ppasta_npasta}, while Fig.~\ref{fig:pcc_ncc} shows the distributions of the same quantities at the interface between the pasta and the core. 
To highlight the importance of a correct treatment of the very low-density region, in both figures the chiral EFT filter on symmetric and pure neutron matter is applied in two density intervals namely $[0.02-0.2]$~fm$^{-3}$ (lower panels) and $[0.1-0.2]$~fm$^{-3}$ (upper panels). 
It can be clearly seen that the application of the constraint from lower density reduces the uncertainties on the crust-core and sphere-pasta transitions, disfavouring the lower and higher values of the transition densities and pressures.
In particular, as already observed in ref.~\cite{dinh2021}, when the EFT constraint is applied from $0.1$~fm$^{-3}$, a considerable fraction of the models predicts low values for both the crust-core and the sphere-pasta transition points.
On the other hand, when the constraint is adopted from $0.02$~fm$^{-3}$, most of the models predicting a crust-core transition below $0.05$~fm$^{-3}$ and very small or even null pasta contribution are excluded from the posterior distribution.

The results of Figs.~\ref{fig:ppasta_npasta} and \ref{fig:pcc_ncc} clearly show that, as far as the crustal properties are concerned, the treatment of the very low-density region has a sizeable impact.
It is interesting to observe that very recently, the authors of ref.~\cite{Shelley2021} conducted a systematic investigation of the composition of the NS inner crust and also underlined the importance of constraining the pure neutron-matter equation of state at subnuclear densities for a reliable description of NS crusts.

The important effect of the behaviour of the functional at densities below $0.1$~fm$^{-3}$ shown by Figs.~\ref{fig:ppasta_npasta} and \ref{fig:pcc_ncc} suggests that, beyond the influence of the $L_{\rm sym}$ parameter that has been advanced by numerous studies \cite{Ducoin2011, Providencia2014}, the high-order parameters might also play a role in the determination of the pasta phase. 
We therefore turn to examine the effect of the chiral EFT filter on the empirical parameters, and the correlation between the transition densities and the behaviour of the energy functional at low density.

\subsection{Low-density equation of state and empirical parameters}
\label{sec:bulk}

\begin{figure*}[!htbp]
\centering
  \includegraphics[width=0.75\linewidth]{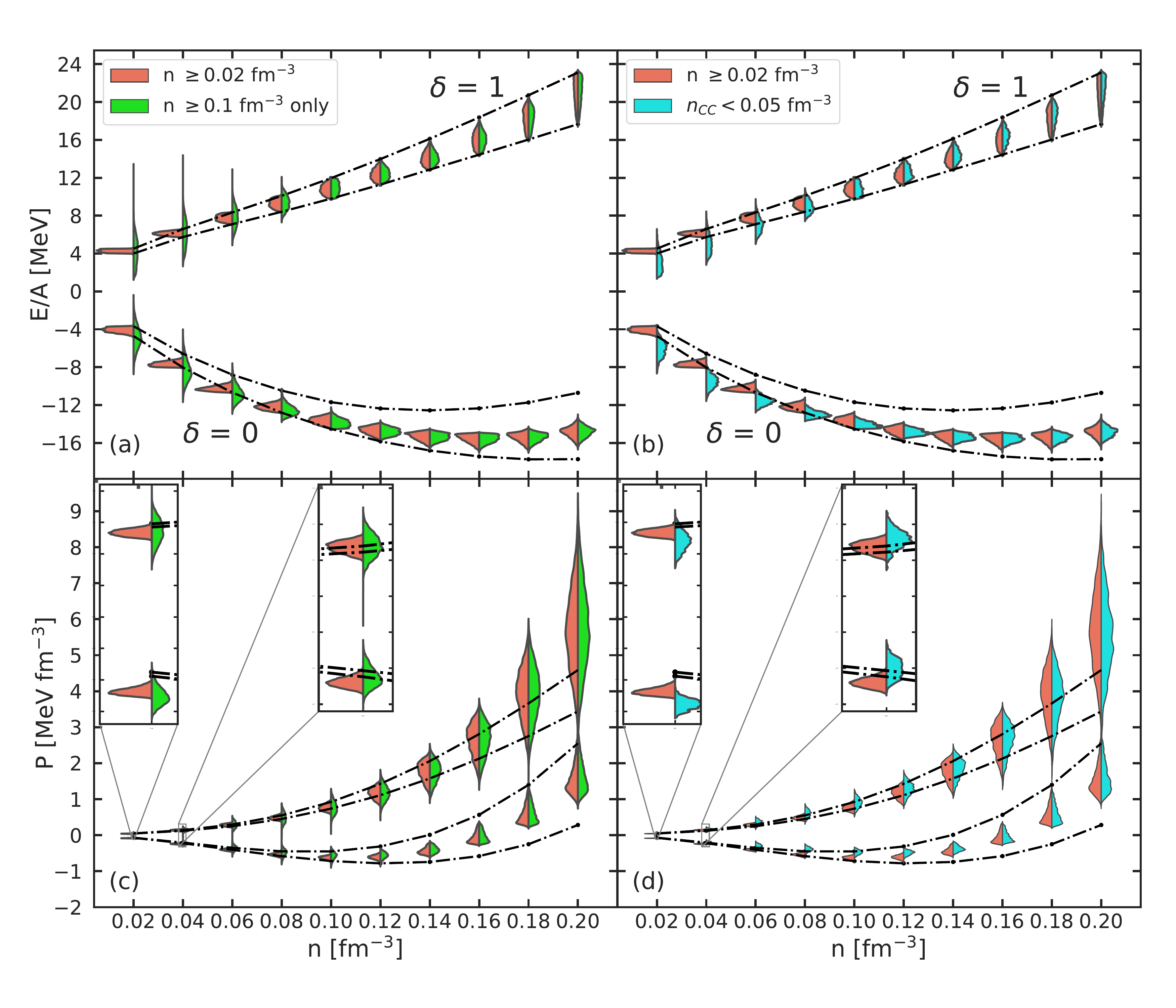}
\caption{Bands of the energy per baryon (top panels) and pressure (bottom panels) of symmetric $(\delta=0)$ and pure neutron matter $(\delta=1)$ as a function of density representing the chiral EFT constraint from \cite{Drischler2016}. 
The probability distributions of models from which EFT constraints are applied from $n \ge 0.02$~fm$^{-3}$ and $n \ge 0.1$~fm$^{-3}$ ($n \ge 0.02$~fm$^{-3}$ and $n \ge 0.1$~fm$^{-3}$ but predicting a crust-core transition $n_{\rm CC} < 0.05$~fm$^{-3}$) are represented as a violin-shape on the left (right) panels. The insets in panels (c) and (d) show a zoom of the low-density part. See text for details.}
\label{fig:eos-nm}       
\end{figure*}

\begin{figure*}
\centering
  \includegraphics[width=0.75\textwidth]{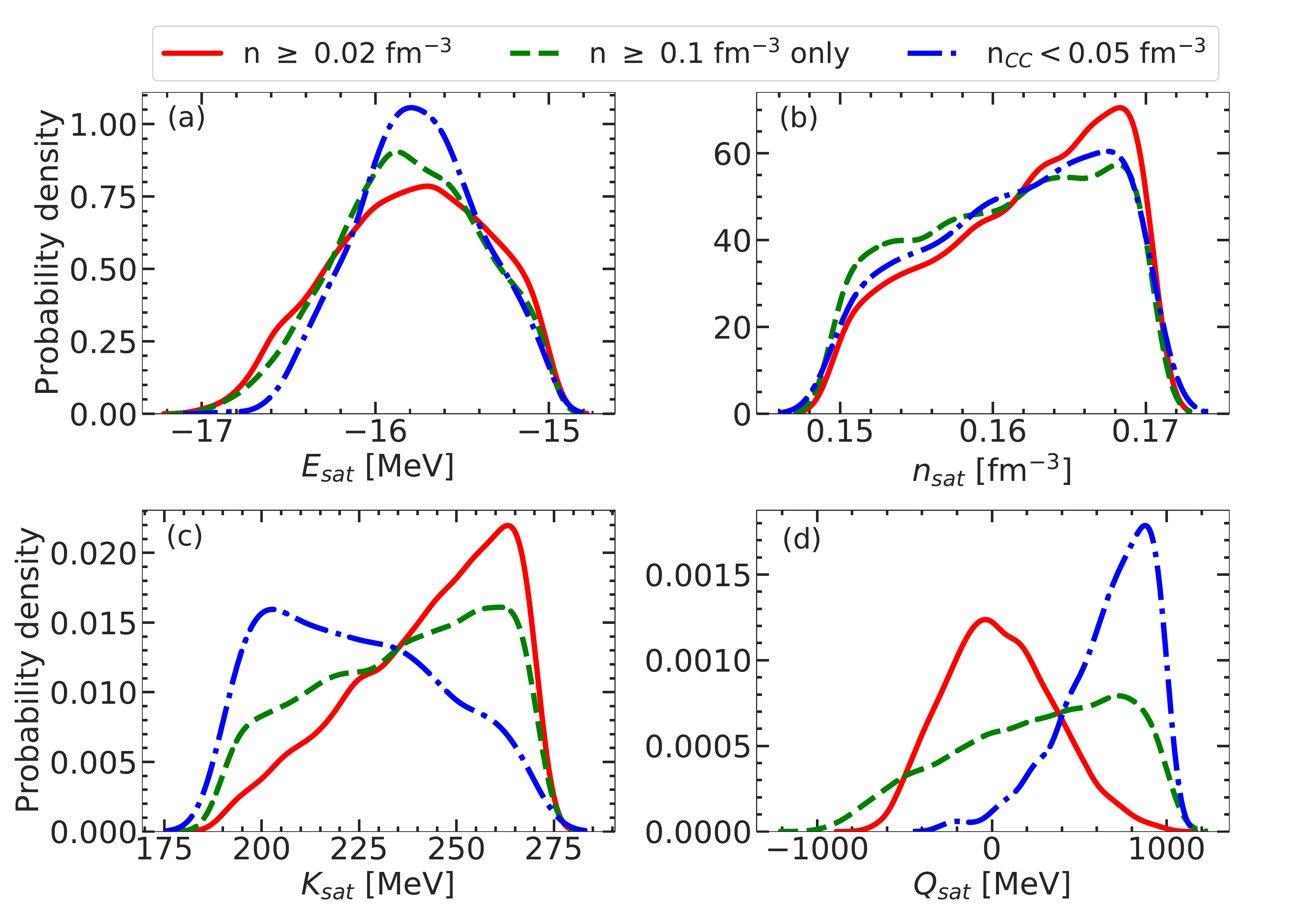}
\caption{Posterior distribution of the isoscalar bulk parameters for models for which the chiral EFT constraint from ref.~\cite{Drischler2016} are applied from $n \ge 0.02$~fm$^{-3}$ (red solid line), $n \ge 0.1$~fm$^{-3}$ (green dashed lines), and from $n \ge 0.1$~fm$^{-3}$ but predicting a crust-core transition $n_{\rm CC} < 0.05$~fm$^{-3}$ (dash-dotted blue line). See text for details.}
\label{fig:is-param}       
\end{figure*}

\begin{figure*}
\centering
  \includegraphics[width=0.75\textwidth]{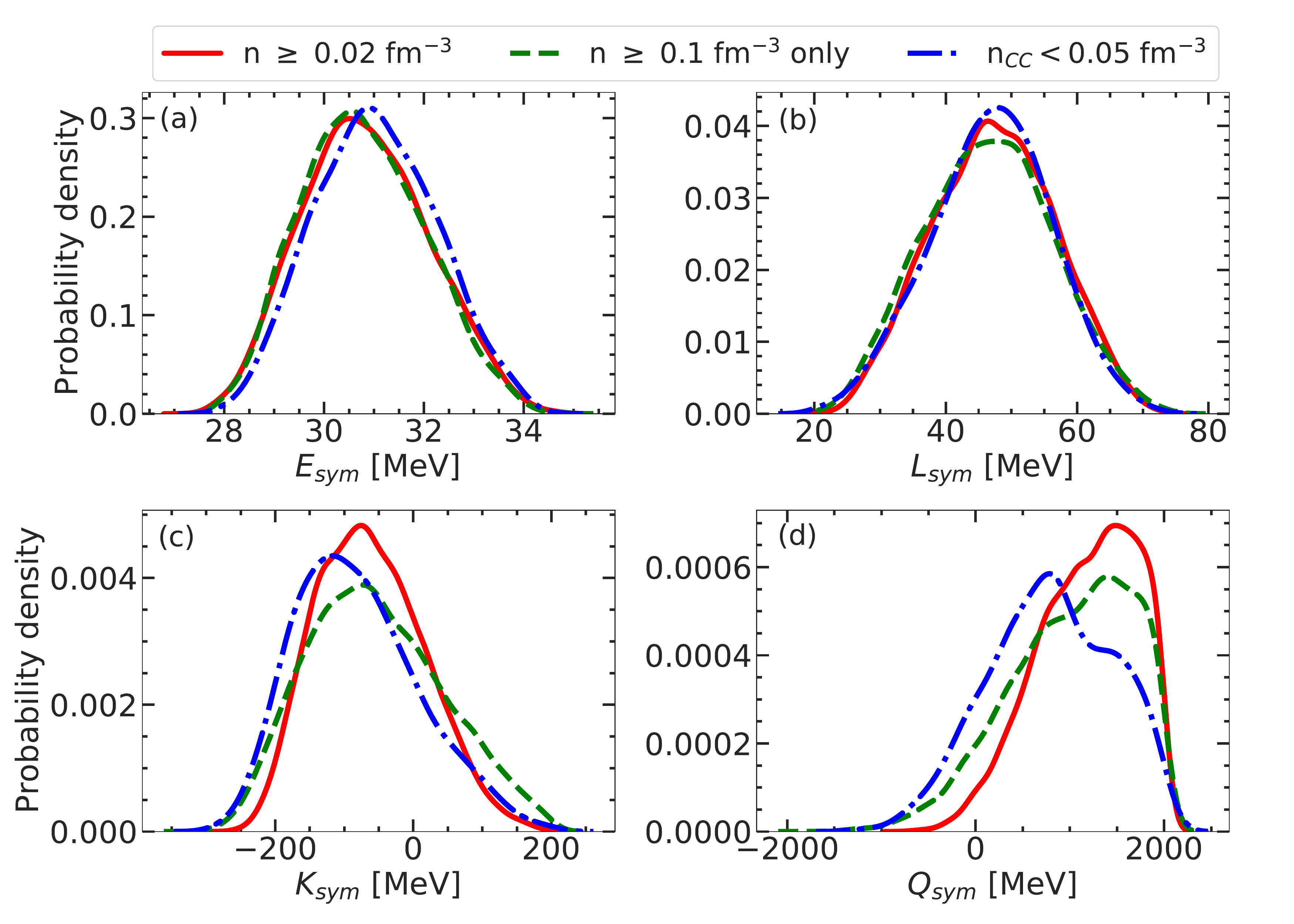}
\caption{Same as in Fig.~\ref{fig:is-param} but for the isovector bulk parameters. See text for details.}
\label{fig:iv-param}       
\end{figure*}

In this section, we discuss the connection between the transition densities and pressures and the behaviour of the low-density energy functional.

Fig.~\ref{fig:eos-nm} shows the distributions of the energy per baryon (upper panels) and pressure (lower panels) of the different models, for symmetric ($\delta=0$) and pure neutron matter ($\delta=1$).
The uncertainty bands from ref.~\cite{Drischler2016} are plotted as black dash-dotted lines, while the violin shapes represent the distributions for the models for which EFT constraints are applied from
(i) $n\ge 0.02$~fm$^{-3}$ (coral shapes on the left part of the density axes in all panels, labelled as ``$n\ge 0.02$~fm$^{-3}$'');
(ii) $n\ge 0.1$~fm$^{-3}$ but not satisfying the constraints in the density range $[0.02-0.1]$~fm$^{-3}$ (green shapes on the right part of the density axes in panels (a) and (c), labelled as ``$n\ge 0.1$~fm$^{-3}$ only'');
(iii) $n\ge 0.1$~fm$^{-3}$ but predicting a crust-core transition lower than $0.05$~fm$^{-3}$ (light blue shapes on the right part of the density axes in panels (b) and (d), labelled as ``$n_{\rm CC} < 0.05$~fm$^{-3}$'').

Comparing panels (a) and (b) of Fig.~\ref{fig:eos-nm}, we can notice that models can violate the EFT constraint both because of a too soft or a too stiff energy behaviour of pure neutron matter at subsaturation densities, but it is the excessive stiffness (corresponding to a too low neutron energy) that leads to abnormally low transition densities. 
In the symmetric matter sector, the mass filter is more constraining than the EFT calculation around saturation and the filter is not effective in that region. 
Going well below saturation, where the mass constraint becomes ineffective, we can observe that the low transition densities are associated to an overbinding of symmetric matter.
Looking at the lower panels, we can see that the low transition densities are globally associated to higher pressures in the sub-saturation region. 
However, a non trivial effect is observed at extremely low densities. 
Indeed we can see that in this regime (right inset in panel (c)) the effect of the filter is to narrow the distribution of the pressure, without sensibly modifying its shape. 
As a result, going to even lower densities (left inset in panel (c)) the pressure is systematically underestimated, showing that the zero-density limit is not correctly reached.
This underlines the fact that the zero-density limit as imposed by ab-initio considerations is not correctly modelled by phenomenological functionals \cite{Yang2016,Grasso2017}. 
From panel (d) we can see that the violation of the EFT predictions is particularly important in the functionals leading to abnormally low transition densities. 
These observations lead us to expect that further improvement in the predictions of the pasta properties might be obtained if the low-density behaviour will be enforced in the functional through the Yang-Lee expansion following refs.~\cite{Yang2016,Grasso2017}.

We now examine the impact of this low-density filter on the bulk parameters, whose distributions are plotted in Figs.~\ref{fig:is-param} and \ref{fig:iv-param} for the isoscalar and isovector parameters, respectively.
We can see that no strong impact is observed on the $E_{\rm sat}$ and $n_{\rm sat}$ isoscalar parameter distributions (see panels (a) and (b) in Fig.~\ref{fig:is-param}), nor on the isovector parameter ones (see Fig.~\ref{fig:iv-param}), except a slight shift of the distributions towards higher values of $K_{\rm sym}$ and $Q_{\rm sym}$ for models filtered from $n\ge 0.02$~fm$^{-3}$.
On the other hand, the low-density filter has a sizeable effect on the higher-order isoscalar parameters $K_{\rm sat}$ and $Q_{\rm sat}$ (see panels (c) and (d) in Fig.~\ref{fig:is-param}).
Indeed, models filtered from $n\ge 0.02$~fm$^{-3}$ (red solid lines) have a more peaked distributions on higher (lower) $K_{\rm sat}$ ($Q_{\rm sat}$) with respect to models predicting $n_{\rm CC} \le 0.05$~fm$^{-3}$ (dot-dashed blue lines).
This behaviour can also explain why models yielding low crust-core transition also have lower energy per baryon and higher pressure at sub-saturation density (see Fig.~\ref{fig:eos-nm}).
Indeed, considering only the lower-order terms in the expansion, Eq.~(\ref{eq:etot}), gives
\begin{eqnarray}
e(n,\delta) &\approx& E_{\rm sat} + \frac{1}{2} K_{\rm sat} x^2 + \frac{1}{6} Q_{\rm sat} x^3 \nonumber \\
  &+& \delta^2 \left( E_{\rm sym} + L_{\rm sym} x + \frac{1}{2} K_{\rm sym} x^2 + \frac{1}{6} Q_{\rm sym} x^3 \right) \ ,
\end{eqnarray}
and, for the pressure,
\begin{eqnarray}
P(n,\delta) &\approx& \frac{n_{\rm sat}}{3} (1+3x)^2 \left[  K_{\rm sat} x + \frac{1}{2} Q_{\rm sat} x^2 \right. \nonumber \\
& +& \delta^2 \left. \left( L_{\rm sym} + K_{\rm sym} x  + \frac{1}{2} Q_{\rm sym} x^2 \right) \right] \ .
\end{eqnarray}
Therefore, roughly speaking, lower average values of $K_{\rm sat}$ and higher values of $Q_{\rm sat}$ result in lower energy per baryon and higher pressure for $x<0$.

From these results, we can infer that the low-energy part of the functional, and particularly  the higher-order isoscalar parameters, have a non-negligible impact on a correct estimation of the transition densities.

\subsection{Surface tension}
\label{sec:surf}

\begin{figure}
  \includegraphics[width=0.45\textwidth]{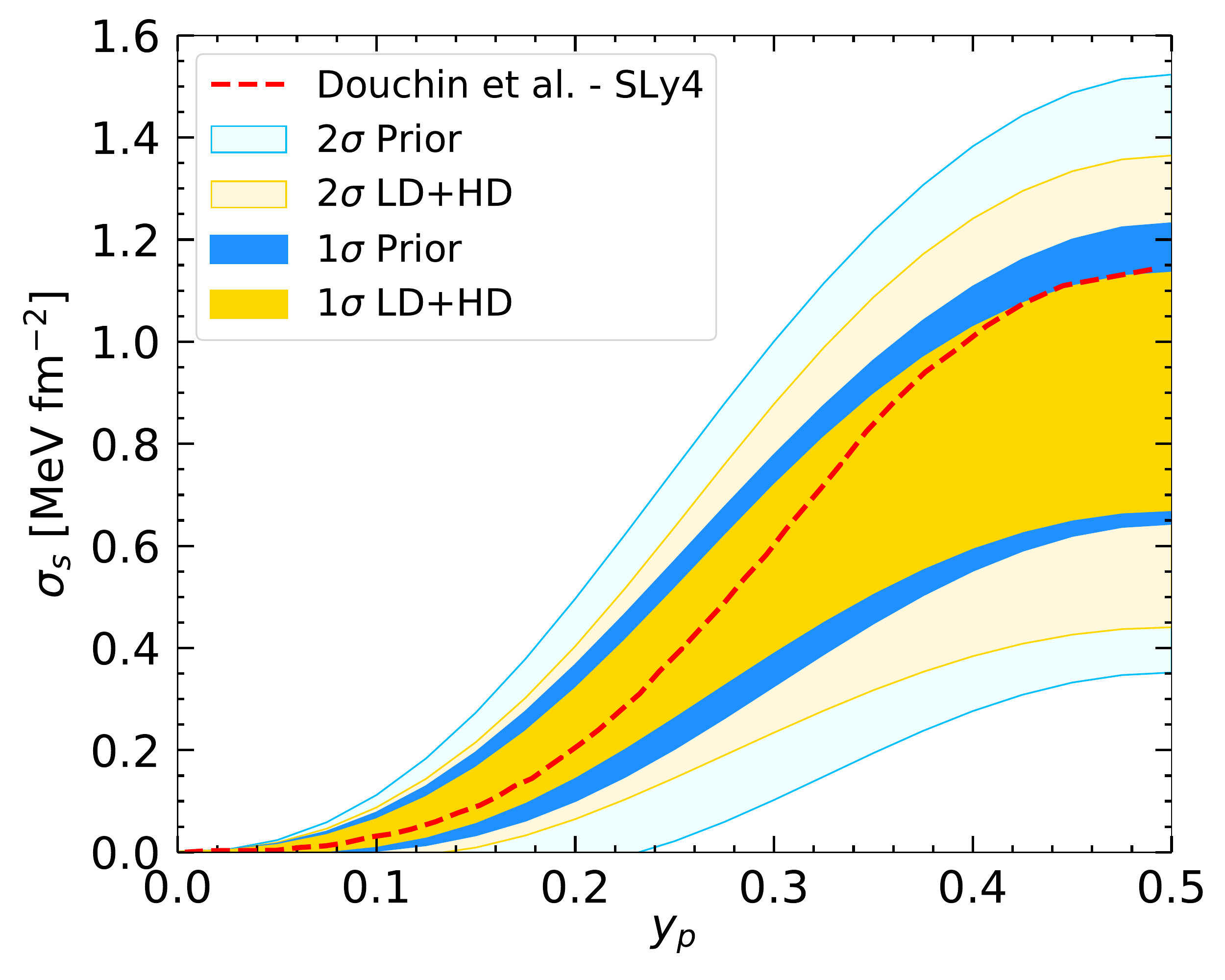}
  \includegraphics[width=0.45\textwidth]{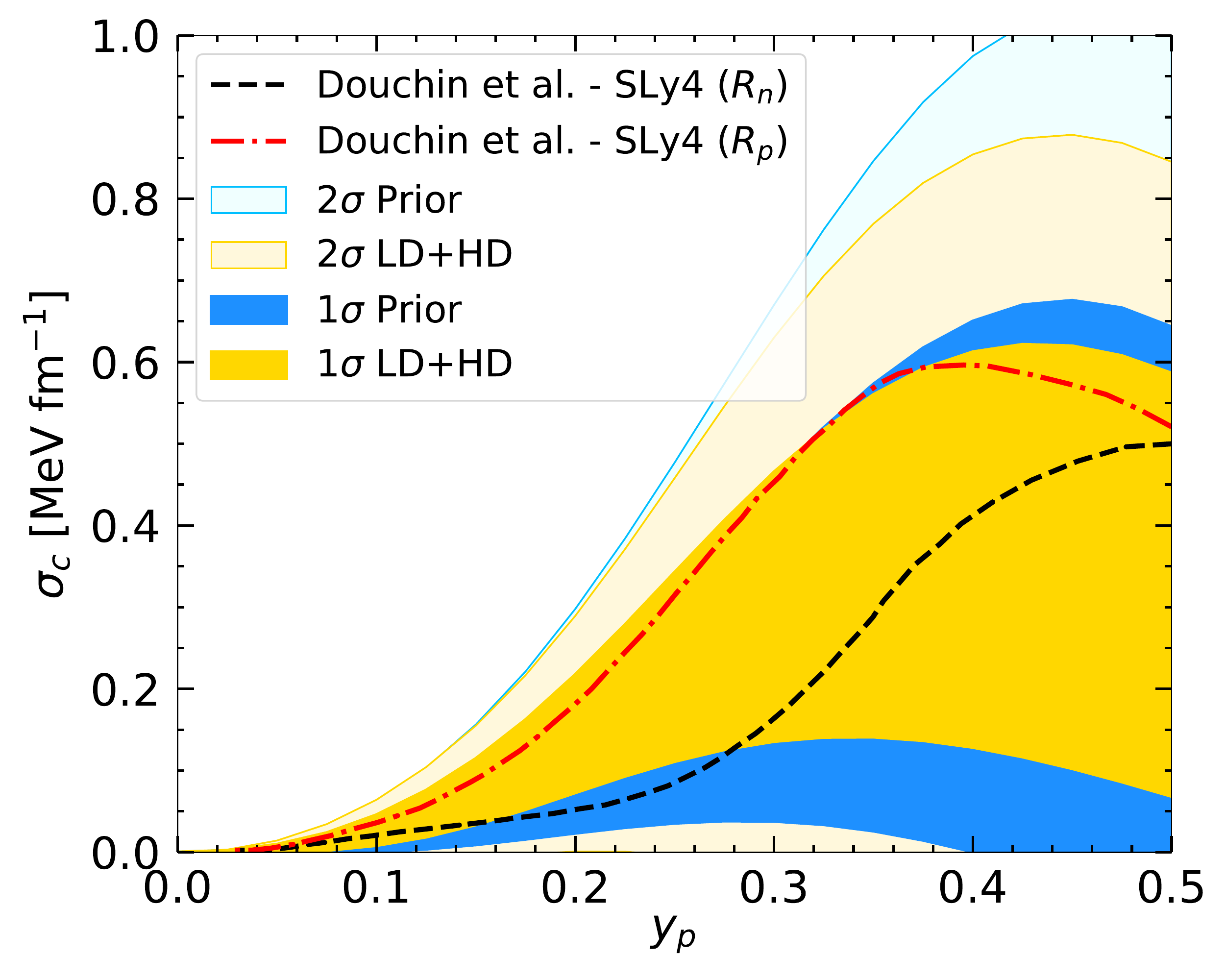}
\caption{$1\sigma$ and $2\sigma$ bands of the surface (top panel) and curvature (lower panel) tensions for the prior (blue bands) and posterior (yellow band) distribution as a function of $y_p=Z/A$. The dashed and dash-dotted lines represent the results from Figs.~3 and 5 of ref.~\cite{Douchin2000}. See text for details.}
\label{fig:sigmasc}       
\end{figure}

\begin{figure}
  \includegraphics[width=0.45\textwidth]{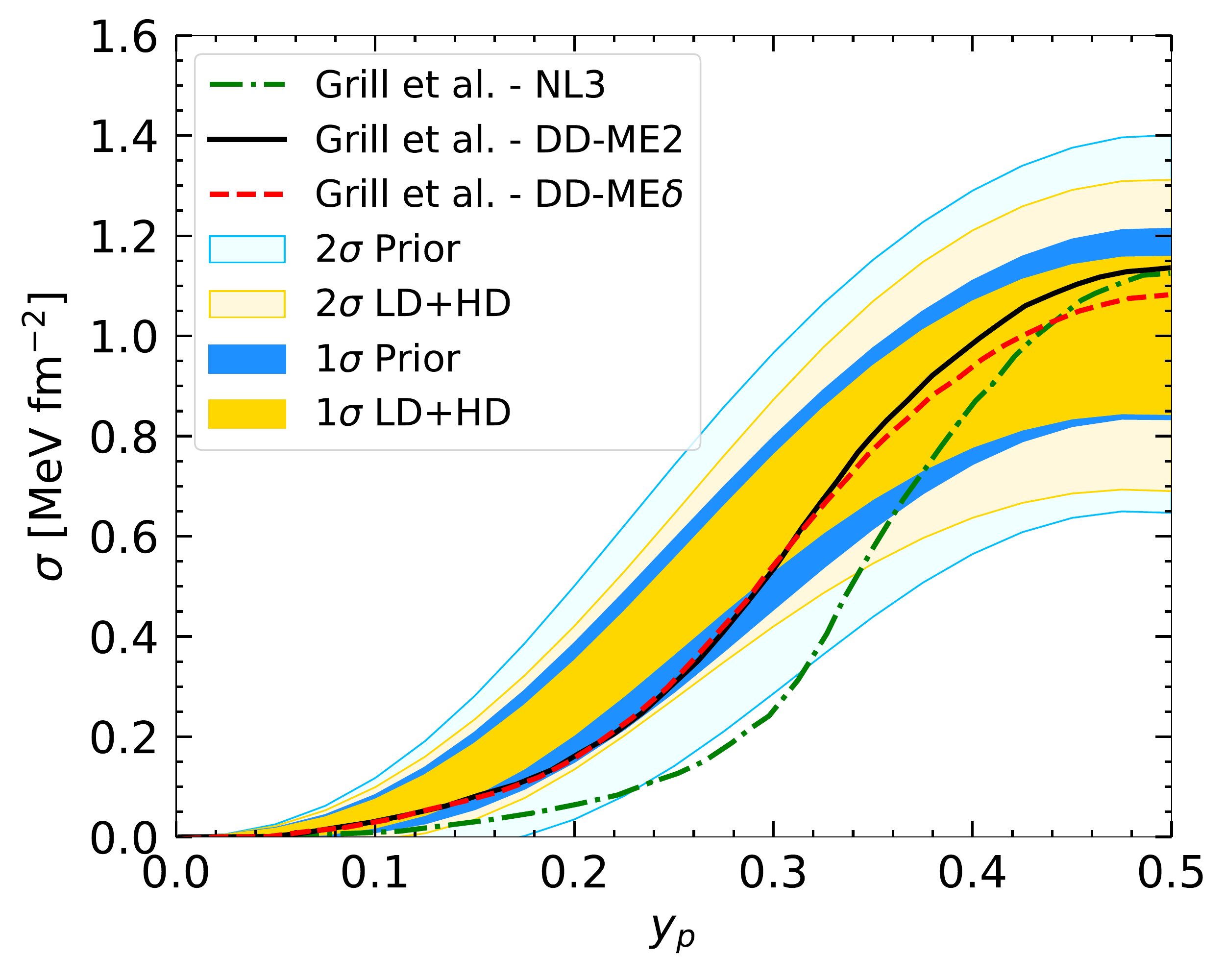}
\caption{$1\sigma$ and $2\sigma$ bands of the total surface tensions for the prior (blue bands) and posterior (yellow band) distribution as a function of $y_p=Z/A$. The solid, dashed, and dash-dotted lines represent the results from Fig.~2 of ref.~\cite{Grill2012} for different models. See text for details.}
\label{fig:sigmatot}       
\end{figure}

\begin{figure}
  \includegraphics[width=0.45\textwidth]{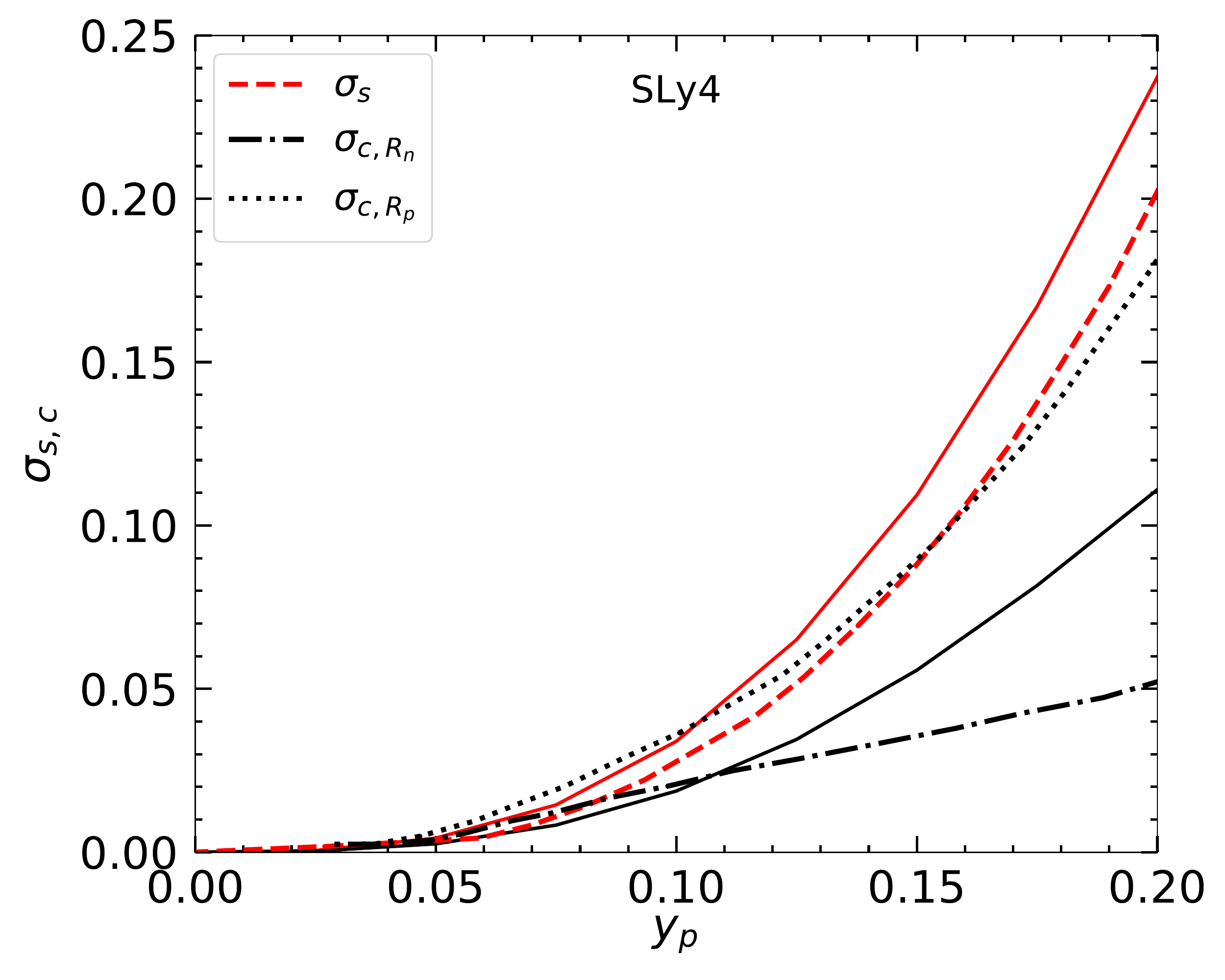}
  \includegraphics[width=0.45\textwidth]{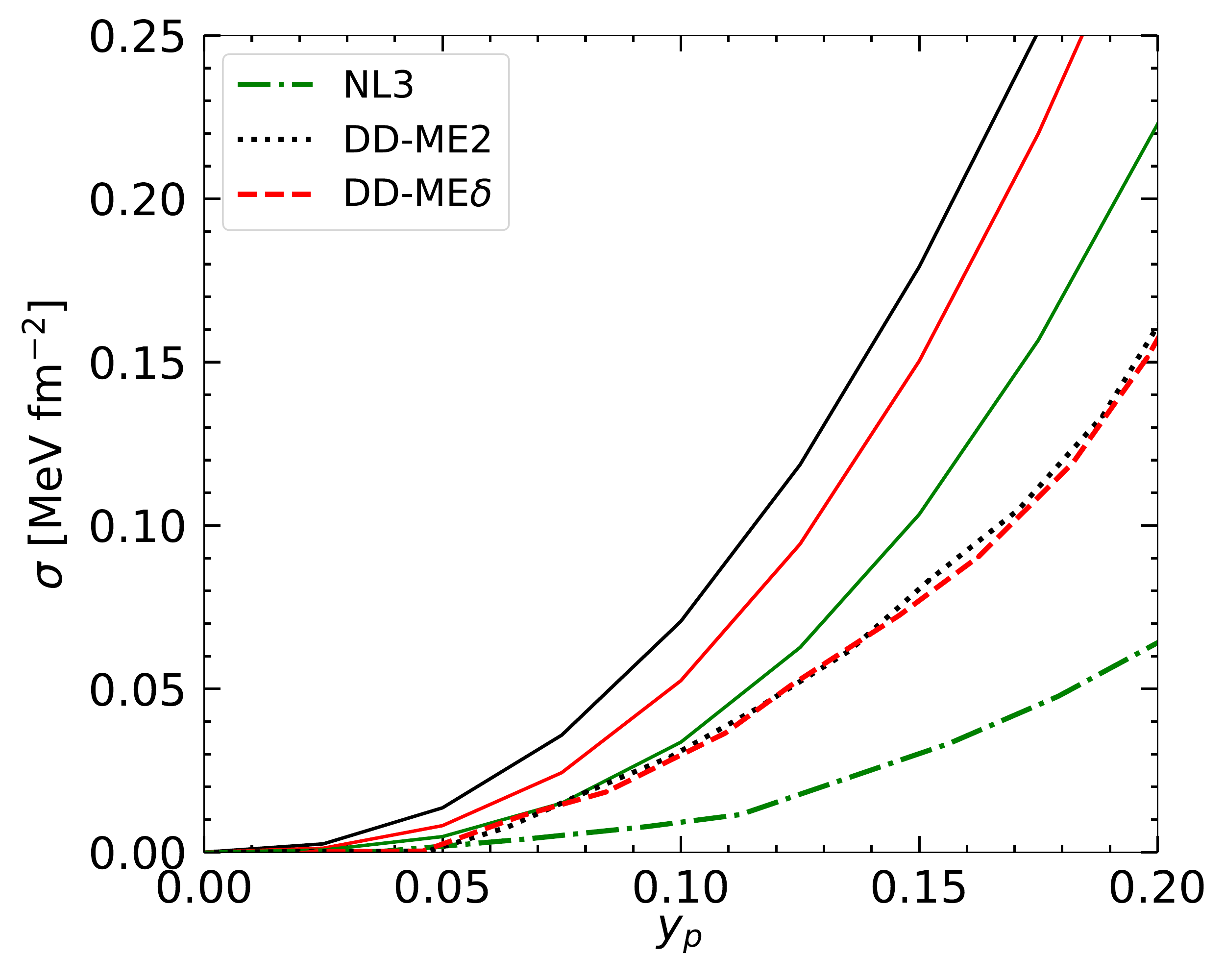}
\caption{Top panel: surface (red lines) and curvature (black lines) tensions as a function of $y_p=Z/A$ calculated for the meta-model SLy4 (solid lines); results from Figs.~3 and 5 of ref.~\cite{Douchin2000} for two reference nuclear surfaces (neutron and proton) are shown as dashed, dotted, and dot-dashed lines.
Bottom panel: total surface tensions as a function of $y_p=Z/A$ for some selected relativistic meta-models (solid lines); results from Fig.~2 of ref.~\cite{Grill2012} are shown as dotted, dashed, and dot-dashed lines. See text for details.}
\label{fig:sigmacomp}       
\end{figure}

We now examine the effect of the finite-size part to the functional, particularly the surface contribution.
In ref.~\cite{dinh2021}, we have shown that the surface parameters are more influential than the bulk ones in the determination of the pasta-phase observables, specifically the fractional pasta radius and moment of inertia with respect to the total radius and moment of inertia of the whole crust.
Moreover, we have underlined the importance of a consistent calculation of the surface and bulk parameters for a reliable evaluation of the uncertainties of the pasta properties.

In our study, the properties of the bulk functional are essentially constrained by the EFT calculation, while independent  surface parameters are introduced, correlated to the bulk properties via the constraint on the reproduction of the nuclear masses. 
As a consequence, the uncertainties in the bulk parameters induce an uncertainty in the surface properties.  
This is consistent with the parameter fitting protocol of Skyrme interactions, for which surface properties are governed by extra gradient terms, with respect to the ones associated to the bulk behaviour. 
In the case of relativistic functionals, however, surface properties emerge naturally from the field equations and cannot be independently varied with respect to the bulk.
One may then wonder if the 5-parameter expressions, Eqs.~(\ref{eq:surface})-(\ref{eq:curvature}), are general enough to account for the different possible behaviours of the surface tension. 
The quality and flexibility of this parametrization was partially verified by Newton et al.~\cite{Newton2013}, who showed that the seminal crust composition of ref.~\cite{bbp} can be indeed reproduced with it, and by Furtado \& Gulminelli \cite{Furtado2020}, who checked that this functional form can very precisely reproduce extended Thomas-Fermi calculations both for terrestrial nuclei and for beyond dripline crustal nuclei in the case of the SLy4 interaction. 
To generalise the discussion, we plot in Figs.~\ref{fig:sigmasc} and \ref{fig:sigmatot} the $1\sigma$ and $2\sigma$ estimation of the surface and curvature tensions as a function of the proton fraction of the denser phase, $y_p = Z/A$.
We can see that the constraint of nuclear masses is not enough to precisely fix the surface tension of symmetric $y_p \approx 0.5$ nuclei, even if they correspond to the quasi totality of the measured masses. 
This can be understood from the degeneracy between the surface and the bulk parameters implied by Eq.~(\ref{eq:mass}), and from the extreme simplicity of the CLD approximation, that does not include shell and pairing effects. 
The absolute uncertainty in the surface tension decreases with decreasing proton fraction, due to the constraint that the surface tension should vanish in pure neutron matter. 
However, this is not the same for the relative uncertainty, that is of the order of $\gtrsim 20\%$ at the typical proton fraction of the clusters in the inner crust, $y_p \lesssim 0.25 $ (see Fig.~\ref{fig:yp}), and even $\gtrsim 60 \%$ for very low proton fractions $y_p \lesssim 0.1$, much higher than the uncertainty on stable $y_p = 0.4-0.5$ nuclei that can be accessed in the laboratory.

Figures \ref{fig:sigmasc} and \ref{fig:sigmatot} also indicate that the choice of the surface and curvature tensions employed here is flexible enough to encompass the results of both CLD model calculations of Douchin et al.~\cite{Douchin2000} with the SLy4 functional (see Fig.~\ref{fig:sigmasc}), and the Thomas-Fermi results from Grill et al.~\cite{Grill2012} obtained with different relativistic mean-field models (see Fig.~\ref{fig:sigmatot}).
In particular, the bottom panel of Fig.~\ref{fig:sigmasc} shows that, for both choices of the reference surface for the curvature tension in ref.~\cite{Douchin2000}, namely the neutron or proton radius (labelled ``$R_n$'' and ``$R_p$'' in the figure, respectively), the results of Douchin et al. \cite{Douchin2000} are within our $1\sigma$ prior distributions.
As for the relativistic mean-field models explored in ref.~\cite{Grill2012}, for which only a comparison of the total surface tension is possible, we can notice that only the NL3 functional is not in agreement with our prior distribution at $1\sigma$, and is marginally compatible with the $2\sigma$ distribution.
Incidentally, the latter functional was also shown to be in disagreement with microscopic calculations of pure neutron matter and with constraints inferred from nuclear-physics experiments on symmetric matter (see Fig.~16 in ref.~\cite{Oertel2017} and Fig.~6.5 in ref.~\cite{burfan2018}).

Despite the flexibility of the expression for the surface tension, model dependence still remains, as it can be seen from Fig.~\ref{fig:sigmacomp}, where we display a zoom of the surface tensions for values of $y_p$ of main interest for this study (see Fig.~\ref{fig:yp}).
In the top panel, we plot the surface (red lines) and curvature (black lines) tensions for the meta-model SLy4 (solid lines) and from the CLD calculations of Douchin et al.~\cite{Douchin2000}.
For the curvature tension, two reference surfaces are considered in ref.~\cite{Douchin2000}, namely the neutron radius $R_n$ and the proton radius $R_p$, while in our case the surface is given by the radius $r_N$, see Eq.~(\ref{eq:interface}).
In the bottom panel, we draw the total surface tension for the meta-models NL3, DD-ME2, and DD-ME$\delta$ (solid green, black, and red lines, respectively), in comparison with the results from the Thomas-Fermi calculations by Grill et al.~\cite{Grill2012} (dash-dotted, dotted, and dashed lines).
Difference up to about a factor of 4 can be seen between our predictions and those in the literature;  this can be essentially attributed to the different protocols to fix the surface parameters. 

This discussion shows that, even within a specific bulk functional, the determination of the surface properties is far from being straightforward. 
For this reason, we estimate that it is very important to take into account the uncertainties in the surface properties within a complete Bayesian analysis, if we want to get realistic error bars in the prediction of crustal properties.

\subsection{Correlations}

\label{sec:corr}
\begin{figure*}
\centering
  \includegraphics[width=0.75\linewidth]{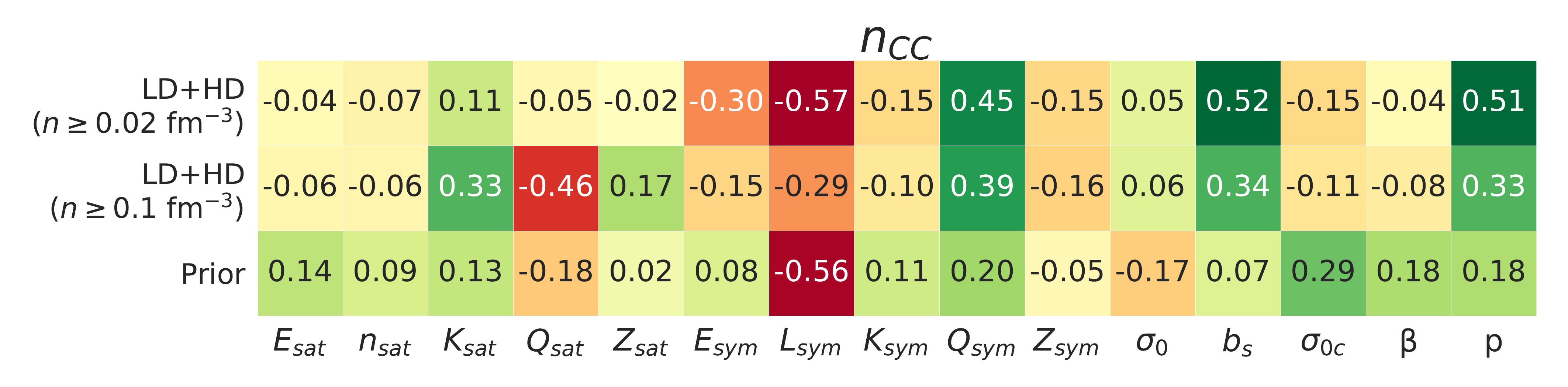}
  \includegraphics[width=0.75\linewidth]{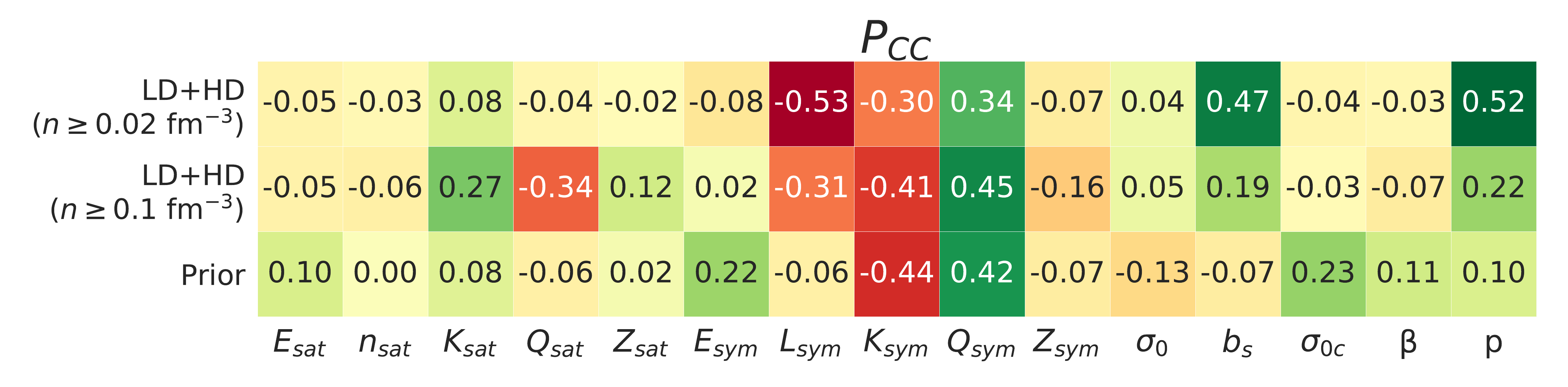}
  \includegraphics[width=0.75\linewidth]{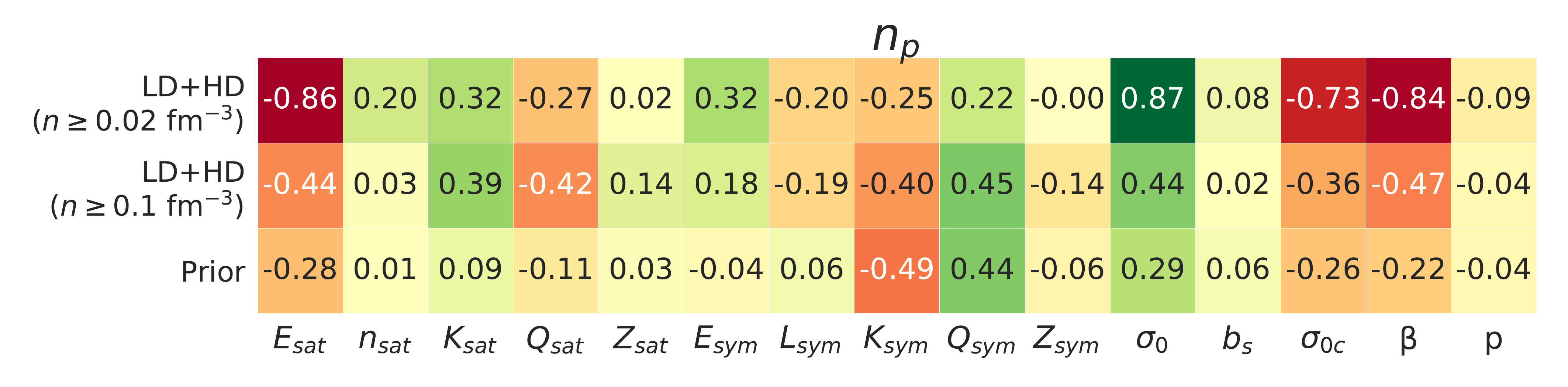}
  \includegraphics[width=0.75\linewidth]{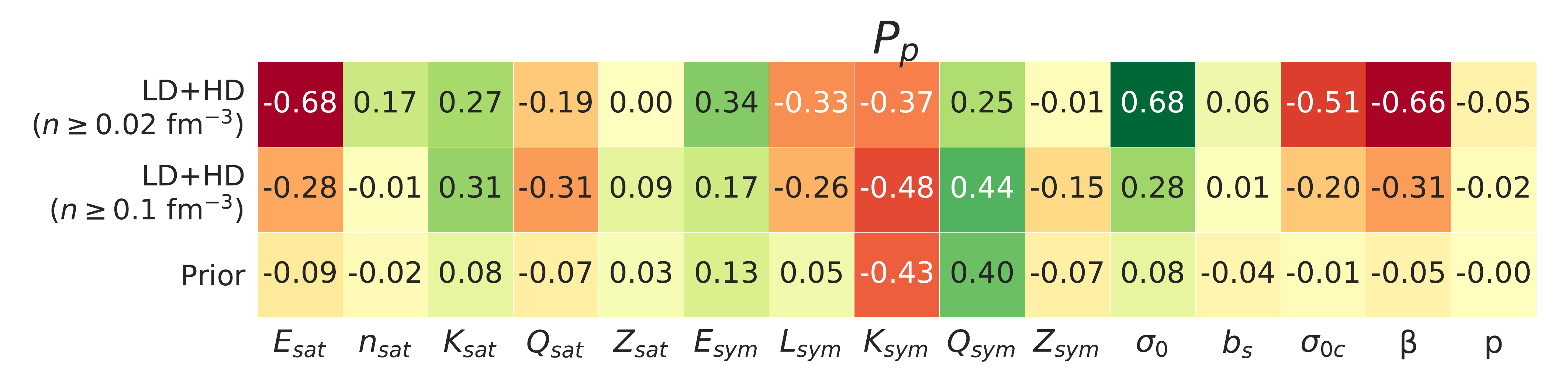}
\caption{Pearson correlations between the crust-core and sphere-cylinder transition density and pressure and the bulk, surface, and curvature parameters.
Two different density intervals for the application of the chiral EFT constraint are considered. See text for details.}
\label{fig:correl}       
\end{figure*}

In the previous sections we have analysed the distributions of the different nuclear parameters that enter in the modelling of the nuclear crust. 
To assess their relative importance in the determination of the pasta properties, we turn to examine the Pearson linear correlation coefficients between the transition observables and the bulk and surface parameters.

In particular, we show in Fig.~\ref{fig:correl} the correlation coefficient for the crust-core transition density and pressure, $n_{\rm CC}$ and $P_{\rm CC}$, and the density and pressure of the transition between spherical and non-spherical configurations, $n_p$ and $P_p$.
For each quantity, we display the correlations for the prior, as well as for the posterior distributions when both low-density (LD) and high-density (HD) filters are accounted for.
Lines labelled as $n \ge 0.02$~fm$^{-3}$ ($n \ge 0.1$~fm$^{-3}$) refer to posteriors for which models have been constrained by EFT calculations in the range $[0.02-0.2]$ ~fm$^{-3}$ ($[0.1-0.2]$~fm$^{-3}$).
As for the crust-core transition, we can see that when no constraints are considered (``prior''), a correlation of the crust-core transition density (pressure) with $L_{\rm sym}$ ($K_{\rm sym}$ and $Q_{\rm sym}$) is noticed, as already pointed out in refs.~\cite{Ducoin2011, Carreau2019}.
A (slight) correlation with the second and third derivatives of the symmetry energy, $K_{\rm sym}$ and $Q_{\rm sym}$, is also observed in the prior for the pasta transition.
However, only when the physical constraints are applied, further interesting correlations start to appear, particularly with respect to the surface parameters when the models are filtered from $n \ge 0.02$~fm$^{-3}$.
Also, a light correlation with $Q_{\rm sat}$ emerges for models filtered from $n \ge 0.1$~fm$^{-3}$, while the correlation with the symmetry parameters is preserved.
As for the pasta transition density and pressure, we can see that only when the EFT constraint is applied from very low density, more significant correlations emerge.
Indeed, otherwise, compensations among the different terms in the functional can occur, thus blurring the correlations.
The most relevant bulk parameters seem to be the energy at saturation, $E_{\rm sat}$, and to a less extent the higher-order derivatives of the symmetry energy $K_{\rm sym}$ and $Q_{\rm sym}$.
From Fig.~\ref{fig:correl}, we can also observe that both the crust-core transition density and pressure and the transition between spherical and pasta configuration are correlated with the surface parameters, particularly when the low-density EFT constraint is enforced from $n \ge 0.02$~fm$^{-3}$. 
In particular, the strong correlation of the crust-pasta transition with $E_{\rm sat}$ might be understood from the important correlation imposed by the mass constraint between $E_{\rm sat}$ and the parameters governing the surface tension at moderate isospin ($\sigma_0$, $\sigma_{0,c}$, $\beta$).
As for the crust-core transition, we can see that it is mainly correlated with the surface parameters, namely $b_s$ and $p$.
Such correlation was already pointed out by Carreau et al.~\cite{Carreau2019}, who observed that, when the low-density filter was considered and the $p$ parameter was allowed to vary, the transition point was correlated to the surface properties and the isovector surface tension was the dominant parameter determining the crust-core transition.
It is encouraging to observe that the fully unconstrained parameter $p$ is not influential at all in the determination
of the pasta transition point (see the last two panels in Fig.~\ref{fig:correl}). 
As far as this latter is concerned, the most important parameters are instead the curvature parameters, namely $\sigma_{0,c}$ and $\beta$, together with $\sigma_0$, that are at least in principle more accessible from experiments, if a more refined model of nuclear mass is employed in the future. 

\section{Conclusions}
\label{sec:concl}

 \begin{table*}
	\caption{Average values and standard deviations of bulk parameters obtained when the EFT constraint is applied in the range $[0.02 - 0.2]$~fm$^{-3}$ and $[0.1 - 0.2]$~fm$^{-3}$.}
	\label{tab:param-bulk}      
	\centering
	\setlength{\tabcolsep}{2pt}
	\begin{tabular}{lcccccccccc}	
		\hline\noalign{\smallskip}
		Filter                 &$E_{\rm sat}$  & $n_{\rm sat}$ & $K_{\rm sat}$ & $Q_{\rm sat}$ & $Z_{\rm sat}$ & $E_{\rm sym}$  &	$L_{\rm sym}$  &$K_{\rm sym}$& $Q_{\rm sym}$ & 	$Z_{\rm sym}$ \\
		                       &[MeV]          & [fm$^{-3}$]   & [MeV]         & [MeV]         & [MeV]         & [MeV]          &   [MeV]          &[MeV]        & [MeV]         &  [MeV]\\
		\noalign{\smallskip}\hline\noalign{\smallskip}
	     $n \ge 0.02$ fm$^{-3}$& -15.80 $\pm$ 0.44  & 0.162 $\pm$ 0.006 & 243 $\pm$ 20 & 46 $\pm$ 308 & 1034 $\pm$ 1192& 30.83 $\pm$ 1.26& 47.3 $\pm$ 9.2& -62 $\pm$ 77& 1186 $\pm$ 521 & 309 $\pm$ 2730\\
	     $n \ge 0.1 $ fm$^{-3}$& -15.81 $\pm$ 0.41  & 0.161 $\pm$ 0.006 & 236 $\pm$ 23 & 265$\pm$ 472 & 801  $\pm$ 1399& 30.78 $\pm$ 1.23& 46.5 $\pm$ 9.7& -55 $\pm$ 98& 1012 $\pm$ 654 & 517 $\pm$ 2738\\
		\noalign{\smallskip}\hline
	\end{tabular}
\end{table*}

\begin{table*}
	\caption{Same as in Table~\ref{tab:param-bulk} but for the surface parameters.}
	\label{tab:param-surf}      
	\centering
	\begin{tabular}{lccccc}	
		\hline\noalign{\smallskip}
		Filter                & $\sigma_0$     & $b_s$  & $\sigma_{0,c}$  & $\beta$  & $p$\\
							  & [MeV fm$^{-2}$]&        & [MeV fm$^{-1}$]&          &  \\
		\noalign{\smallskip}\hline\noalign{\smallskip}
		$n \ge 0.02$ fm$^{-3}$&0.90268$\pm$0.23105 & 32.37735 $\pm$ 30.54776 &0.15956 $\pm$ 0.03050&0.83368 $\pm$ 0.25541& 3.00 $\pm$ 0.58\\
		$n \ge 0.1 $ fm$^{-3}$&0.90822$\pm$0.21394 & 34.26680 $\pm$ 31.19063 &0.15554 $\pm$ 0.02904&0.82919 $\pm$ 0.24080& 3.03 $\pm$ 0.57\\
		\noalign{\smallskip}\hline
	\end{tabular}
\end{table*}

In this work, we have studied the properties of the pasta phases in cold catalysed NSs, within a CLD with parameters adjusted on experimental masses or theoretical calculations.
We have employed different nuclear (meta-)models to study the model dependence of the results, as well as a statistical analysis to quantitatively estimate the uncertainties in the predictions.
All the considered models predict the existence of pasta phases (spheres, rods, slabs, tubes, and eventually bubbles), but the transition densities among the geometries are strongly model dependent. 

To understand the origin of the model dependence and pin down the most relevant parameters, we have performed a full Bayesian analysis by largely exploring the bulk and surface parameter space, and imposing constraints both from nuclear physics and astrophysics, the most important ones coming from ab-initio nuclear theory, and from the experimental knowledge of nuclear mass. 
The chiral EFT calculations are seen to considerably constrain the low-order empirical parameters, but important uncertainties persist in the high-order ones, particularly $Q_{\rm sym}$ and $Q_{\rm sat}$. 
This is seen in Table~\ref{tab:param-bulk} that summarises the average values and standard deviation of the parameters for our posterior distributions when the EFT filter is applied in the range $[0.02 - 0.2]$~fm$^{-3}$ (first line) and $[0.1 - 0.2]$~fm$^{-3}$ (second line).
Even if the relative uncertainty in the low-order parameters is relatively small, their important correlation with the surface parameters imposed by the nuclear mass constraint induces an important dispersion in the estimation of the surface tension (see Table~\ref{tab:param-surf}), that in turn is highly influential in the determination of the pasta properties. 
This uncertainty might contribute to explain the important model dependence observed in the pasta modelling in the literature.

A correlation study reveals that the density and pressure of the crust-pasta and pasta-core transition are strongly correlated both to surface and to bulk parameters. 
Concerning the crust-core transition, we confirm that the most influential bulk parameters are $L_{\rm sym}$,  $K_{\rm sym}$ and $Q_{\rm sym}$. 
As already stressed in ref.~\cite{Carreau2019}, the isospin dependence of the surface tension, that is not strongly constrained by nuclear masses, also plays an important role.
Concerning the location of the core-pasta transition, once the bulk functional is optimised within the results of chiral EFT for homogeneous neutron matter, the surface tension at moderate isospin turns out to be the dominant ingredient. 
This quantity is in principle accessible from the measurement of nuclear mass, but a more sophisticated mass model must be introduced to reduce the present uncertainties.

 The modelling of the NS crust presented here is suitable for applications in astrophysical simulations and/or data analyses.
The NS equation of state and composition obtained with such models, provided with their nuclear parameters and relative error bars, could be tabulated and included in open-access databases such as the CompOSE \cite{compose} database for direct use in astrophysical applications to obtain predictions of NS observables with controlled uncertainties.


\begin{acknowledgements}
This work has been partially supported by the IN2P3 Master Project NewMAC and the CNRS International Research Project (IRP) “Origine des \'el\'ements lourds dans l’univers: Astres Compacts et Nucl\'eosynth\`ese (ACNu)”. 
\end{acknowledgements}



\end{document}